\documentclass[twocolumn,superscriptaddress]{revtex4}

\usepackage{epsfig,wrapfig}
\usepackage{amssymb,amsfonts,amsmath}
\usepackage{afterpage}

\renewcommand{\k}{\kappa}

\renewcommand{\t}{\tau}
\newcommand{\E}{\mathbb{E}}

\newcommand{\ud}{\,\mathrm{d}}

\newcommand{\Li}{{\rm Li}}

\newcommand{\si}{Supporting Information}
\newcommand{\mm}{Materials \& Methods}
\newcommand{\beq}{\begin{equation}}
\newcommand{\eeq}{\end{equation}}

\begin{document}

\title{Hierarchy and Extremes in Selections from Pools of Randomized Proteins}

\author{S\'ebastien Boyer} 
\affiliation{Laboratoire Interdisciplinaire de Physique, CNRS \& Universit\'e Grenoble Alpes, Grenoble, France}

\author{Dipanwita Biswas} 
\affiliation{Laboratoire Interdisciplinaire de Physique, CNRS \& Universit\'e Grenoble Alpes, Grenoble, France}

\author{Ananda Kumar Soshee} 
\affiliation{Laboratoire Interdisciplinaire de Physique, CNRS \& Universit\'e Grenoble Alpes, Grenoble, France}

\author{Natale Scaramozzino} 
\affiliation{Laboratoire Interdisciplinaire de Physique, CNRS \& Universit\'e Grenoble Alpes, Grenoble, France}

\author{Cl\'ement Nizak} 
\affiliation{Laboratoire de Biochimie, CNRS \& ESPCI-ParisTech, Paris, France}

\author{Olivier Rivoire} 
\affiliation{Laboratoire Interdisciplinaire de Physique, CNRS \& Universit\'e Grenoble Alpes, Grenoble, France}

\begin{abstract}
Variation and selection are the core principles of Darwinian evolution, yet quantitatively relating the diversity of a population to its capacity to respond to selection is challenging. Here, we examine this problem at a molecular level in the context of populations of partially randomized proteins selected for binding to well-defined targets. We built several minimal protein libraries, screened them {\it in vitro} by phage display and analyzed their response to selection by high-throughput sequencing. A statistical analysis of the results reveals two main findings: first, libraries with same sequence diversity but built around different ``frameworks'' typically have vastly different responses; second, the distribution of responses within a library follows a simple scaling law. We show how an elementary probabilistic model based on extreme value theory rationalizes these findings. Our results have implications for designing synthetic protein libraries, for estimating the density of functional biomolecules in sequence space, for characterizing diversity in natural populations and for experimentally investigating the concept of evolvability, or potential for future evolution.
\end{abstract}

\maketitle

Diversity is the fuel of evolution by natural selection but translating this concept into quantitative measurements is not straightforward~\cite{magurran2013measuring}. A simple count of the number of different individuals in a population for instance fails to account for the very different responses to selection that two populations with same number of different individuals may elicit. The problem is already acute at the molecular scale where it also takes a very practical form: libraries of diverse proteins are routinely screened as a way to identify biomolecules of interest (binders, catalysts,\dots) and a proper ``diversity'' is critical for success~\cite{Zhao:1997we,Wong:2006uw}. But beyond a general agreement that maximizing the number of different elements is desirable, there is no general rule for engineering and comparing diversity in these libraries.

A common design of many protein libraries is to concentrate variations at one or a few variable parts located around a fixed ``framework'', which is shared by all members of the library~\cite{Zhao:1997we,Wong:2006uw}. The natural design of antibody repertoires, the pools of immune proteins with potential to recognize nearly every molecular target, follows this pattern. Most of sequence variations in antibodies are indeed concentrated at a few loops extending from a common structural scaffold~\cite{Padlan:1994wq}. This design has inspired the conception of artificial protein libraries built on frameworks other than the immunoglobulin fold~\cite{Urvoas:2012ef}.

Here, we propose a new approach to quantitatively characterize the selective potential of molecular libraries. To develop this approach, we designed and screened 24 synthetic protein libraries with identical sequence variations but different frameworks and analyzed their response to well-defined selective pressures by high-throughput sequencing. Between libraries, we find that selective potentials vary widely and define a hierarchy of frameworks. Within libraries, we find that selective potentials exhibit a simple scaling law, characterized by few parameters. The essence of these results is captured by an elementary probabilistic model based on extreme value theory (EVT). This leads us to propose a new measure of the selective potential of a population that overcomes the shortcomings of previously proposed measures of diversity.

\section{Methods}

\subsection{Library design} 
\begin{figure}[t]
\centering
\includegraphics[width=.6\linewidth]{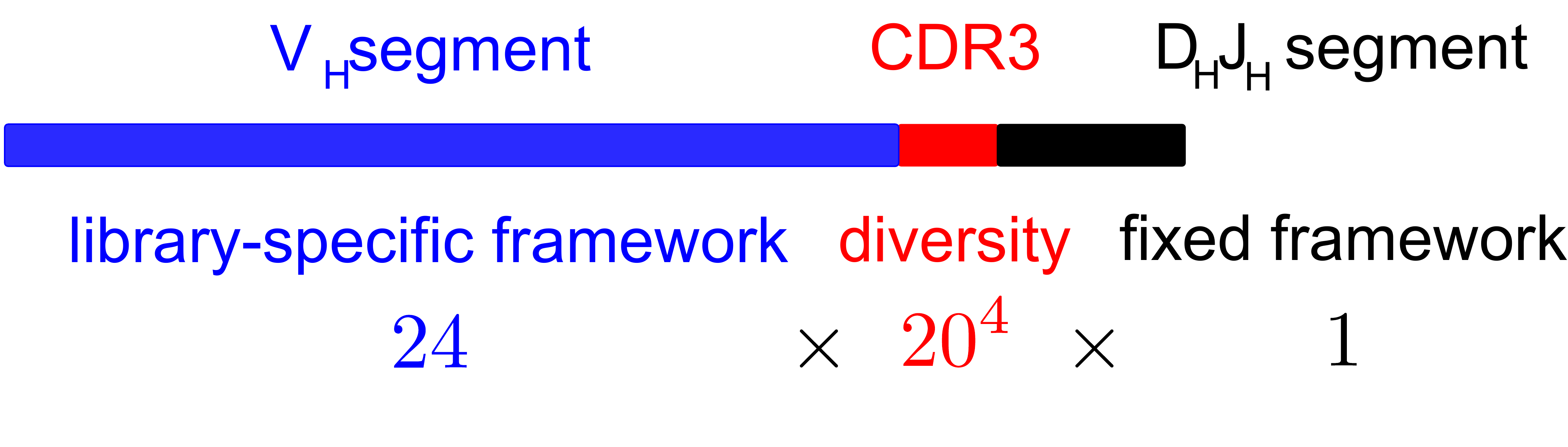}
\caption{Library design -- We designed a total of 24 libraries with distinct frameworks and identical sequence diversity consisting of all $20^4=1.6\times 10^5$ combinations of the 20 natural amino acids at 4 consecutive positions. The design follows the natural design of the variable ($V$) region of the heavy ($H$) chain of antibodies, which is assembled by joining three gene segments, the variable ($V_H$), diversity ($D_H$) and joining ($J_H$) segments. The library-specific parts of the frameworks (in blue) are from natural $V_H$ and diversity is introduced at the third complementarity determining region (CDR3, in red), at the junction between $V_H$ and $D_HJ_H$, a part of the sequence critical for specific binding to antigens; the $D_H$ and $J_H$ segments (in black) are common to all libraries.\label{fig:lib}}
\end{figure}

We built 24 minimal libraries with different frameworks but identical sequence diversity (\mm, Figures~\ref{fig:lib} and S1). Twenty frameworks consist of single-domain antibodies taken from natural heavy-chain genes of diverse origins ($V_H$ fragments), typically sharing $40\%$ of their amino acids (Figure S2); they originate from maturated antibodies, which are mutated relative to their germline form, except for the S1 framework, which comes from a germline (na\" ive) antibody. Three additional frameworks are more closely related and correspond to the germline and two maturated forms of the same human antibody, with the maturated frameworks sharing 65\% and 85\% sequence identity with the germline. Finally, one framework consists exclusively of glycines to serve as a control. Diversity is limited to four consecutive amino acids at the complementarity determining region 3 (CDR3), the part of antibody sequences most critical for specificity~\cite{Xu:2000vq}. Structurally, the CDR3 forms one of three loops that define the binding pocket of a $V_H$ domain~\cite{Padlan:1994wq}; in our design, the two other loops  (CDR1 and 2) are thus part of the framework. Our libraries are minimal on two accounts: the framework consists of a single domain of $\sim 100$ amino acids and the total diversity is $20^4=1.6\times 10^5$ -- all combinations of the 20 natural amino acids at the 4 varied sites. For comparison, the most commonly used antibody libraries consist of two domains ($V_H$ and $V_L$) and have $>10^8$ variants, with variation introduced at different CDRs~\cite{Hoogenboom:2005fo}. Libraries based on $V_H$ only are, however, known to be effective~\cite{Ward:1989bs}. ``Minimalist libraries'' have also been built by restricting the alphabet of amino acids at the variables sites but contained $>10^{10}$ variants~\cite{Fellouse:2004ez,Fellouse:2005ku,Fellouse:2007cv}. One of the simplest libraries demonstrated so far, built on a synthetic scaffold, still contained $>10^6$ variants randomly sampled from a much larger pool of potential sequences~\cite{Fisher:2011fy}.

\subsection{Selection} We screened our libraries by phage display for binding to one of two targets, a neutral synthetic polymer, poly-vinylpyrrolidone (PVP), and a short DNA loop of 9 nucleotides (\mm). Two previous studies established the capacity of antibody phage display to select binders for these targets~\cite{Soshee:2014ku,Modi:2013eb}. Phage display is a standard high-throughput screening technique~\cite{Smith:1997vn}. It is based on the fusion of each antibody sequence to the sequence of the pIII surface protein of the filamentous bacteriophage M13, a natural virus of the bacterium {\it E.~coli} with the shape of a 1~$\mu$m long and 10~nm wide cylinder~\cite{Smith:1997vn}. The engineered phage encapsulates the DNA sequence of an antibody and displays the corresponding polypeptide at its surface. Populations of up to $10^{14}$ phages displaying a total diversity of up to $10^{10}$ different antibodies can thus be manipulated. A round of selection consists in retrieving the phages bound either to the bottom of a plate where the PVP target is attached or to magnetic beads where the DNA target is coated. It is followed by a round of amplification achieved by infecting bacteria with the selected phages. We performed experiments where each sequence is initially present in $\sim 10^4$ copies and where targets are provided in excess. Starting either from a single library (single framework) or from a mixture of different libraries, three rounds of selection/amplification were performed. Although the enrichment of some of the sequences is intended to reflect binding to the specified targets, other factors may contribute, such as sequence-specific differences in amplification. In our experiments, such non target-specific selective factors can be detected but are non-dominant (\si). Our analysis and its interpretation, however, do not rely on the precise nature of the selective pressure.

\subsection{High-throughput sequencing} We sequenced samples of $10^6-10^7$ sequences at different rounds of selection by Illumina Miseq pair-ended high-throughput sequencing~\cite{Shendure:2008jh}. The results give us an estimation of the relative frequencies $f_i^t$ of each sequence $i$ in the population at each round $t=0,1,2,3$. In estimating these frequencies, we take into account both sequencing and sampling errors (\mm).

We define the selectivity to a target of each sequence $i$ as its probability $s_i$ to pass a round of selection. This selectivity is inferred up to a multiplicative factor $a$ as the ratio of the frequency of the sequence before and after selection~\cite{Fowler:2010gt}:
\beq\label{eq:s_i}
s_i=a\ \frac{f_i^{t}}{f_i^{t-1}}.
\eeq
The unknown multiplicative constant $a$ reflects our lack of quantitative control over the rate of amplification of the sequences. In our calculations, we arbitrarily fix $a$ so that $\sum_i s_i=1$; we explain below how our conclusions depend on this choice. We compare the frequencies between rounds $t=3$ and $t-1=2$, where sequences with highest selectivities are best represented.

Previous studies have applied next-generation sequencing to the outcome of phage display screens as a way to identify a large number of binders~\cite{dias2009next,Ravn:2010kq} but have not investigated the distribution of the relative selectivities of these binders. 

\subsection{Reproducibility and specificity}  Several observations based on the frequencies and amino-acid patterns of the sequences in populations under selection validate our experimental approach. (1)~Screening the same library against the same target in separate experiments yields reproducible frequencies $f_i^t$ at the last round $t=3$ (Figure~S3). (2)~Screening the same library against different targets yields target-specific amino-acid patterns (Figures~S4). (3)~Screening two libraries against the same target yields library-specific amino-acid patterns (Figure~S4). Taken together, these results show that enrichment of some of the sequences is reproducible and arises from selection for specific binding to the targets. We note that one feature of our experiments is critical for reproducibility: the initial populations maximize degeneracy (the number of copies of each sequence) rather than diversity (the number of distinct sequences).

\section{Results}

\subsection{Hierarchy between libraries}

\begin{figure}[t]
\centering
\includegraphics[width=\linewidth]{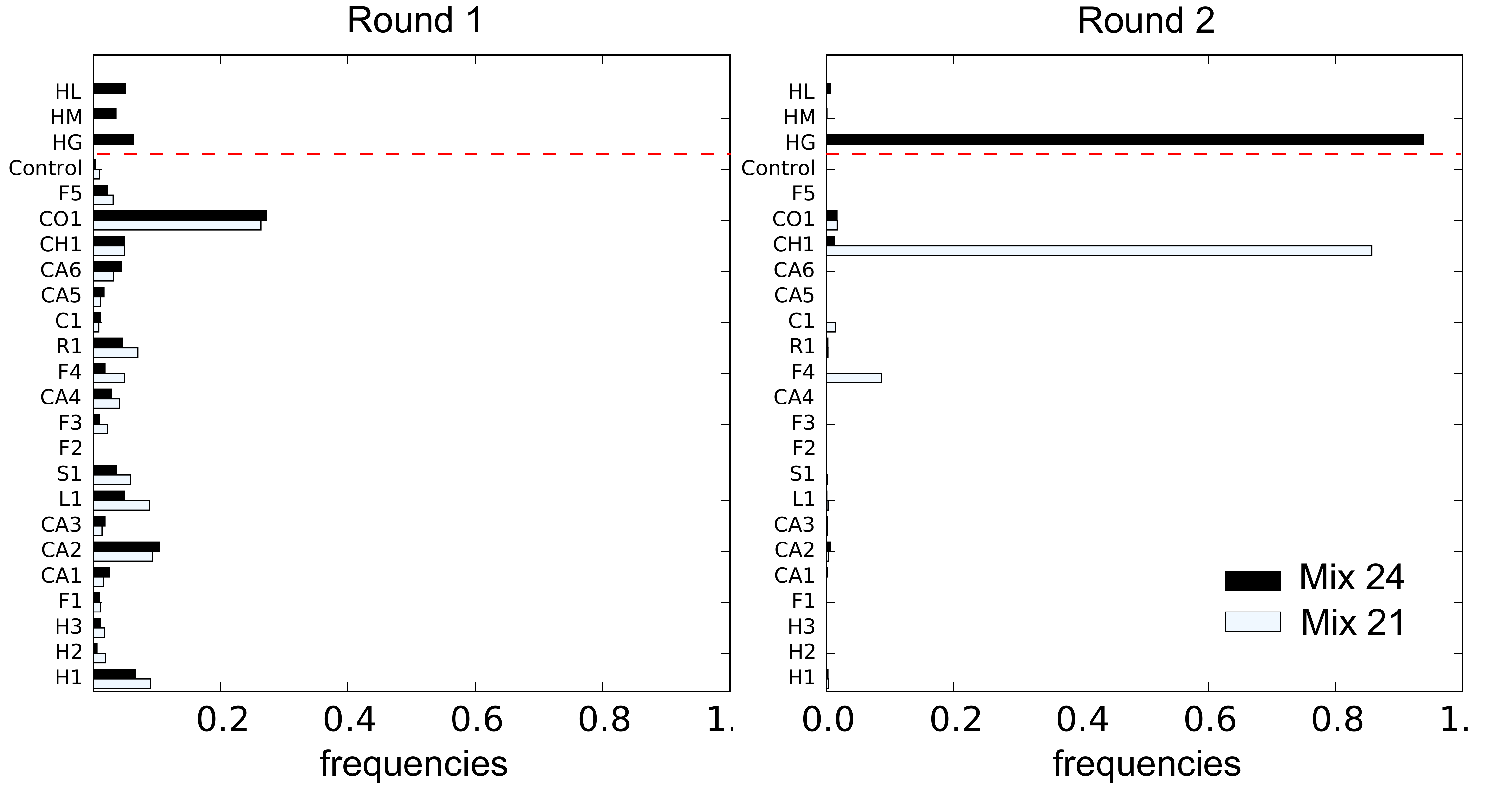}
\caption{Hierarchy between libraries -- Frequencies of the different libraries in two successive rounds of selection against the DNA target. Black bars report selection of all 24 libraries and white bars selection of a subset of 21 libraries, excluding the 3 libraries above the red dotted line. At the second round (right), the population is enriched in sequences from one particular library, the HG library, in contrast to what is observed at the first round (left). The subset of 21 libraries excludes the library dominating the mixture of all 24 libraries, which leads another library, the CH1 library, to dominate. Within the two libraries, a diversity of CDR3 are selected (Figures~\ref{fig:plaw}B and \ref{fig:plaw}D), with different patterns of amino acids (Figure S4). Enrichment from the other libraries can also be observed when they are screened in isolation (\si).\label{fig:inter}}
\end{figure}

To compare the selective potentials of libraries built around different frameworks, we performed experiments in which the initial population of sequences consists of a mixture of libraries with distinct frameworks -- a meta-library. The results of these experiments reveal a striking hierarchy. Diverse members of a same library, i.e., sequences sharing a common framework, typically dominate. When repeating the experiment with an initial mixture of libraries that excludes the dominating library, another library dominates (Figure \ref{fig:inter}). Libraries not selected when mixed with other libraries nevertheless do contain sequences with detectable selectivities, as shown by screening them in isolation (Figure S4). These results are not explained by uneven representations of the libraries in the initial population or by framework-specific differences during amplification (Figure~S5).

Differences in frameworks are thus generally more significant than differences between variable parts, even though these parts are clearly under selection for binding (Figures~\ref{fig:plaw}B and \ref{fig:plaw}D). This result may not be surprising for very dissimilar frameworks, but our frameworks are all expected to share the same structural fold and some frameworks have few sequence differences. In particular, the dominating framework when selecting the mixture of all 24 libraries against the DNA target (Figure \ref{fig:inter}) is a germline human $V_H$ framework, which dominates two libraries built on frameworks derived from it by affinity maturations, which share respectively 65\% and 85\% of their amino acids. The observed hierarchy is target dependent: different frameworks dominate when screening the meta-library against different targets. Remarkably, when screening the 24 libraries against the PVP target (Figure S6), the dominating framework is the only other germline framework of the mixture (the S1 framework). As noted previously, differences between frameworks also appear in the patterns of amino acids that are selected at the level of CDR3s (Figures S4C-E).
 
\subsection{Scaling within libraries}

\begin{figure}[t]
\centering
\includegraphics[width=.99\linewidth]{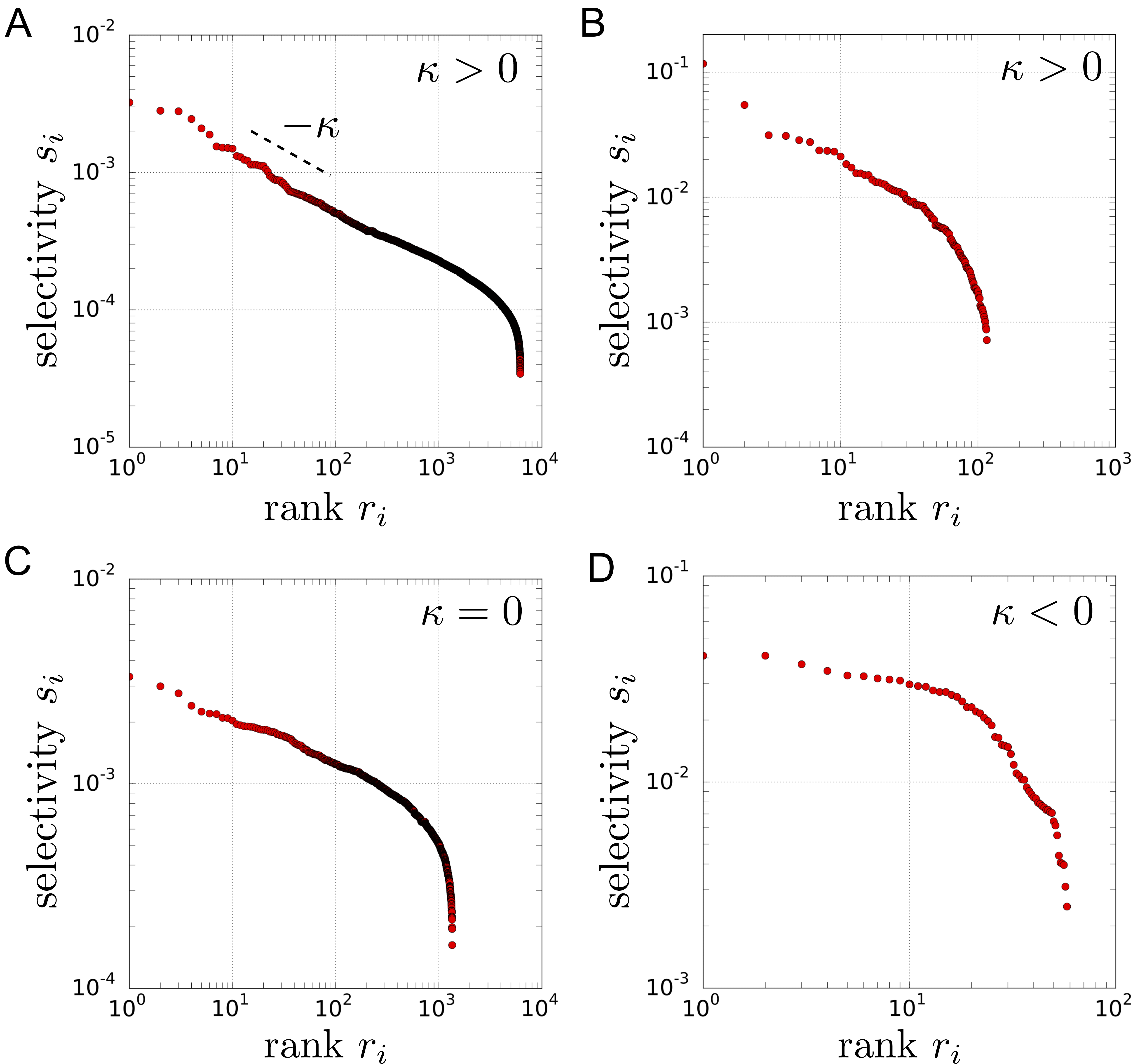}
\caption{Scaling relations within libraries -- The selectivities $s_i$ of the sequences are represented versus their ranks $r_i$ for four experiences differing by the input library and the choice of the target against which it is selected. {\bf A.} S1 library against the PVP target. {\bf B.} HG library against the DNA target. {\bf C.} F3 library against the PVP target. {\bf D.} CH1 library against the DNA target. In A, the distribution of the top $\sim 1000$ sequences follows a power law with exponent $\k\simeq 0.5$. This behavior is consistent with the prediction of extreme value theory (EVT) when the shape parameter is positive, $\k>0$ (see Figure~\ref{fig:stat} for the analysis that justifies this conclusion). Although not obvious from this representation, the data in B is also consistent with EVT when $\k>0$ while the data in C and D are consistent with EVT when, respectively, $\k= 0$ and $\k<0$.\label{fig:plaw}}
\end{figure}

To compare the selectivities of sequences sharing a common framework,  and therefore differing by at most 4 amino acids (Figure \ref{fig:lib}), we rank these sequences in decreasing order of their selectivity $s_i$ and plot these selectivities versus the ranks on a double logarithmic scale -- a representation of the cumulative distribution of selectivities within a library. For several experiments, this representation reveals a power law: if $s(r)$ is the selectivity of the sequence of rank $r$, then, for the sequences with top ranks,
\beq\label{eq:scaling}
s(r)\sim r^{-\kappa}.
\eeq
Figure~\ref{fig:plaw}A shows an example where the exponent is $\kappa\simeq 0.5$. While this power law is observed for several libraries (different frameworks) and selective pressures (different targets), it is not systematic: deviations are often observed for the very top sequences (Figure~\ref{fig:plaw}B) and, for several experiments, a power law cannot be justified (Figure~\ref{fig:plaw}D).

Both the power law and its various deviations can, however, be rationalized under an elementary mathematical model. This model rests on two assumptions. First, it assumes that the selectivity of each sequence in a library is drawn independently at random from a common probability density $\rho(s)$, which may depend on the framework and the target. Second, it assumes that the sequences with top selectivities are in the tail of this probability density.

The model is thus probabilistic even though -- baring out experimental noise -- the experiments have no inherent stochastic element. To the extent that selectivity reflects binding at thermodynamical equilibrium, the selectivity $s_i$ of antibody $i$ is indeed determined by its binding free energy $\Delta G_i$ to the target: $s_i\propto e^{-\Delta G_i/k_BT}$, where $T$ represents the temperature and $k_B$ the Boltzmann constant. The binding free energy $\Delta G_i$ is a physical quantity which, in principle, is fully determined by the sequence of amino acids. In the spirit of applications of random matrix theory to nuclear physics~\cite{mehta1967random}, it may nevertheless be advantageous to discard this microscopic description in favor of a coarser probabilistic description, which treats the selectivities $s_i$ as instances of random variables independently drawn from a common probability density $\rho(s)$. In contrast to nuclear physics, no symmetry constrains $\rho(s)$ {\it a priori} but, if concerned only with the largest $s_i$, results from extreme value theory (EVT), the branch of probability theory dealing with extrema of random variables~\cite{gumbel1958statistics}, do constrain the form of the tail of $\rho(s)$ from which they originate, thus allowing for non-trivial predictions.

EVT indeed indicates that random variables $s$ independently drawn from the tail of a common probability density have themselves a probability density of the form~\cite{coles2001introduction}
\beq\label{eq:gp1}
f_{\k,\t,s^*}(s)=f_\k\left(\frac{s-s^*}{\tau}\right),
\eeq
with $f_{\k}$ necessarily belonging to the generalized Pareto family:
\beq\label{eq:gp2}
f_\k(x)=
 \left\{
      \begin{aligned}
        &(1+\k x)^{-\frac{\k+1}{\k}}\quad &{\rm if\ } \k\neq 0,\\
        &e^{-x}\quad &{\rm if\ } \k= 0,\\
      \end{aligned}
    \right.
\eeq
where the exponential for $\k=0$ is just the continuous limit of $f_\k(x)$ when $\k\to 0$.
Here, $s^*$ represents a threshold above which the tail of $\rho(s)$ is defined, $\tau$ is a scaling factor (which absorbs the undetermined factor $a$ introduced in Eq.~\eqref{eq:s_i}) and $\kappa\geq -1$ is the so-called shape factor (independent of $a$), which defines the universality class to which the distribution of selectivities belongs: the probability densities $\rho(s)$ may differ, but if they are associated with the same $\k$, events drawn form their tails will share similar statistical properties.

As suggested by the notations, when $\kappa>0$, but only when $\kappa>0$, this model predicts that the top ranked sequences follow a power law with exponent $\kappa$, as described by Eq.~\eqref{eq:scaling}. Mathematically, when considering a large number $N$ of samples, the rank $r(s)$ is indeed related to the cumulative distribution of selectivities by
\beq\label{eq:rank}
r(s)\sim N\int_s^\infty \rho(x) \ud x.
\eeq
If $\rho(s)\sim s^{-\frac{\k+1}{\k}}$ for large $s$ as predicted by Eq.~\eqref{eq:gp2} for $\k>0$, we must then have $\int_s^\infty \rho(x) \ud x\sim s^{-1/\k}$ and therefore $r(s)\sim s^{-1/\k}$, which is equivalent to Eq.~\eqref{eq:scaling}. 
In other words, the power law seen in Figure~\ref{fig:plaw}A corresponds to the expected relationship between the rank and the values of random variables drawn from the tail of a probability density when this density belongs to a class associated with $\k>0$. 

\begin{figure}[t]
\centering
\includegraphics[width=\linewidth]{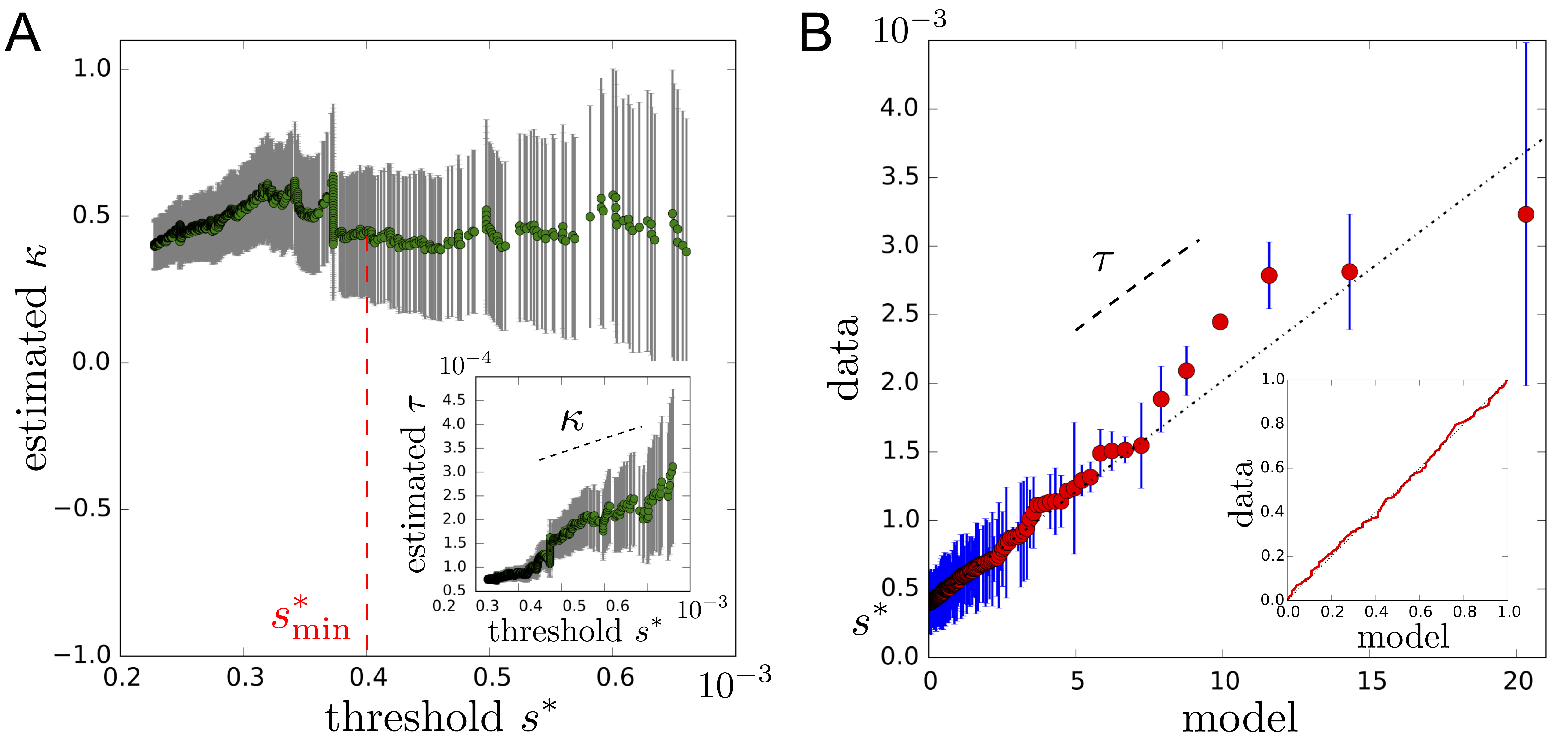}
\caption{Extreme value analysis by the point-over-threshold approach -- {\bf A.} Values of the inferred parameters $\hat\k(s^*)$ from selectivities $s_i>s^*$ as a function of the threshold $s^*$. The inference is made by maximum likelihood and the error-bars indicate 95\% confidence intervals. Inset: similarly for  $\hat\tau(s^*)$, the second parameter of the model, which is estimated jointly to $\k(s^*)$. For $s^*$ sufficiently large, $s^*>s_{\rm min}^*$, $\hat\k(s^*)$ should be constant and  $\hat\tau(s^*)$ increase linearly with slope $\hat\k(s^*)$. This is observed here for $s_{\rm min}^*\simeq 4\times 10^{-4}$ (red dotted line) and leads to $\k=0.45\pm 0.22$ and $\tau=1.6\times 10^{-4}\pm 10^{-5}$; $\k=0$ can be excluded by likelihood ratio test with a p-value $<10^{-4}$. {\bf B.} Quantile-quantile plot representing the data $s_i$ against predictions from the model based on  the inferred value of $\k$ only. A straight line is expected for a good fit with a slope and the y-intercept given by the two other parameters $\tau$ and $s^*$. Inset: probability-probability plot comparing the empirical cumulative distributions from the data to the cumulative distribution from the inferred model, showing an excellent agreement. The data for this figure comes from the selection of the S1 library against the PVP target as in Figure~\ref{fig:plaw}A (see Figures S8-S10 for a similar analysis of the data shown in Figures~\ref{fig:plaw}B-D).\label{fig:stat}}
\end{figure}

To precisely assess the ability of our model to describe all these different cases, we followed the point-over-threshold approach, a standard method in applications of EVT to empirical data~\cite{coles2001introduction}. This approach consists in fitting the data $s_i$ satisfying $s_i>s^*$ by a function of the form $f_{\k,\t,s^*}(s)$ for different values of the threshold $s^*$, and then in estimating whether a threshold $s^*_{\rm min}$ exists such that for $s^*>s^*_{\rm min}$ the inferred parameter $\hat\k(s^*)$ is nearly independent of $s^*$. To apply this method, we inferred the parameters $\hat\k(s^*)$ and $\hat\t(s^*)$ by maximum likelihood from the data $s_i>s^*$ for every value of $s^*$. For the data presented in Figure~\ref{fig:plaw}A, an illustration is provided in Figure~\ref{fig:stat}A, with error-bars indicating 95\% confidence intervals (see \si\ for the analysis of other experiments). In this example, we observe that $\hat\k(s^*)$ becomes nearly constant, of the order of 0.5, for $s^*>s_{\rm min}^*\simeq 4\times 10^{-4}$. The determination of $s_{\rm min}^*$ is performed by visual inspection but any choice of $s^*>s_{\rm min}^*$ should give equivalent results.

Given $s^*>s_{\rm min}^*$ and the associated values of $\k=\hat\k(s^*)$ and $\t=\hat\t(s^*)$ inferred from maximum likelihood, the next step is to estimate whether this best fit is indeed a good fit. The diagnosis is commonly performed visually using probability-probability (P-P) and quantile-quantile (Q-Q) plots~\cite{coles2001introduction}. The P-P plot compares the empirical and modeled cumulative distributions by representing the quantile function $q(s)=r(s)/N$ (the fraction of the data above $s$) against the cumulative $F_{\hat\k,\hat\t,s^*}(s)=\int_0^s f_{\hat\k,\hat\t,s^*}(x)\ud x$. As indicated by Eq.~\eqref{eq:rank}, a straight line $y=x$ is expected if the fit is perfect, which the inset of Figure~\ref{fig:stat}B shows to be nearly the case in this example. The Q-Q plot makes a similar comparison but by representing $s$ against $F^{-1}_{\hat\k,0,0}(q^{-1}(s))$, where $q^{-1}(x)$ represents the value of $s$ above which a fraction $x$ of the data is located. This representation has two advantages over the P-P plot: it relies only on the estimation of $\k$ and it displays more clearly the contribution of the most extreme values. A straight line is expected if the fit is perfect, but this time with a slope $\t$ and a y-intercept $s^*$. The main panel of Figure~\ref{fig:stat}B indicates again a very good fit in the illustrated case.

Performing the same analysis on results of selections of various libraries against various targets, we find that the model is able to describe all the experiments (Figures S8-S12). Different values of $\kappa$ are obtained with differences that are statistically significant (Table S1). In particular, the three cases $\k>0$, $\k=0$ and $\k<0$ are represented.

While many models can lead to a power law~\cite{Mitzenmacher:2004va}, our probabilistic model has the merit of explaining the various deviations from this behavior that the data exhibits. First, when $\k>0$, EVT predicts a power law with exponent $\k$ for the top-sequences but accounts for deviations both for the very top-ranked sequences, which, under the model, may vary widely (Figure~S7), and for sequences of smaller selectivities, where $f_\k$ in Eq.~\eqref{eq:gp2} can provide an excellent fit well beyond the point where the power law applies (e.g., Figures~\ref{fig:stat}B and S8). Second, EVT predicts behaviors differing from a power law if the probability density $\rho(s)$ belongs to an universality class associated with $\kappa\leq 0$, consistently with the results of some of the experiments (e.g., Figures~\ref{fig:plaw}D and S10).

\section{Discussion}

We presented a quantitative analysis of {\it in vitro} selections of multiple libraries of partially randomized proteins with variations limited to four consecutive amino acids. The distribution of selectivities of the top sequences is described by few parameters, with an interpretation provided by an elementary probabilistic model based on extreme value theory (EVT). 

Within a library whose members share a common framework, this distribution is characterized by a shape parameter $\kappa$, which may be either positive, negative or zero. This parameter is independent of the unknown factor $a$ in Eq.~\eqref{eq:s_i} and has several interpretations. For instance, it controls the relative spacing between selectivities: ranking the sequences from best to worst, the expected difference of selectivity between sequences at rank $r$ and $r+1$, $\Delta_r=\E[s_{r}-s_{r+1}]$, satisfies $\Delta_r/\Delta_1\sim r^{-(\k+1)}$, i.e., the larger $\k$, the wider the spread between phenotypes in the library (\si). The shape parameter also provides a statistical answer to the following question: if sampling $N$ sequences yields a top sequence of selectivity $s_1$, what best selectivity $s'_1$ may we expect from sampling $N'>N$ sequences? The difference $\E[s_1'-s_1]$ is a sharply increasing function of $\k$ (\si; Figure S13); as consequence, multiplying by a factor 1000 the number of sequences when $\k=0$ is expected to have same effect as multiplying it by a factor 2 when $\k=0.2$, if starting with $N=10^5$ sequences.

Besides the shape parameter $\k$, the other parameters are the scaling parameter $\tau$, the threshold of selectivities $s^*$ that defines where the tails starts, and the fraction $\phi$ of the data above this threshold (there is some freedom in the choice of $s^*$ on which both $\tau$ and $\phi$ depend: see \si). Within our experimental set-up where the selectivities are determined only up to a multiplicative factor (see Eq.~\eqref{eq:s_i}), the values of $s^*,\phi$ and $\tau$ obtained from different experiments cannot be directly compared, but our selections with mixtures of libraries suggest that $s^*$ varies from library to library on a scale larger than the scale of the differences of selectivity within libraries. All the parameters of the model are found to be both framework and target dependent (Table S1).

Based on these results, we propose these parameters as general descriptors of the selective potential of a population of random variants facing a given selective constraint. In particular, these descriptors could be applied to re-visit the fundamental problem of estimating the density of functional proteins or RNAs in sequence space. Previous studies have estimated this density by counting the number of different sequences enriched in {\it in vitro} selections~\cite{Ellington:1990ct,Keefe:2001fa}. The results of such experiments depend on experimental noise, which sets a lower limit $s_{\rm noise}$ on detectable selectivities. In turn, our approach is dependent only on the library content and the selective pressure, provided $s^*>s_{\rm noise}$.

Power laws are seemingly ubiquitous in distributions of protein features~\cite{Huynen:1998uha,Koonin:2002gn}. Most closely related to our work, the distribution of abundances of distinct antibody sequences in Zebrafishes has been shown to follow a power law with exponent $\alpha\simeq 1$~\cite{Weinstein:2009he,Mora:2010jx}. Only instantaneous frequencies, not selectivities, are accessible in such a case, but, assuming a homogeneous initial distribution of sequences, frequencies and selectivities have same distribution and $\alpha=\k$ if $\k>0$. However, repeating $n$ times the same selection leads to $\alpha=n\k$, which does not account for a stable exponent $\alpha>0$, which may arise in natural repertoires from fluctuating selective pressures~\cite{Desponds:2015ve}. One possible extension of our approach could be to explore this scenario by changing the target between successive rounds of selection.

While many models can be consistent with a power law, our model based on EVT covers without additional assumption the deviations from a power law observed in the data; in particular, it can fit the data over a larger range of selectivities and account for non-power-law behaviors. This is, however, not the first application of EVT to the description of biological variation: Gillespie first introduced it in models of evolutionary dynamics as a way to constrain the distribution of beneficial effects obtained when mutating a wild-type individual~\cite{Gillespie:1982wx,gillespie1991causes}. He assumed $\k=0$, arguing that this class includes all ``well-behaved'' distributions, among which the exponential, normal, log-normal and gamma distributions. Mathematical models for the distribution of affinities in combinatorial molecular libraries have also proposed that it should have universal features but only considered distributions in the exponential class $\k=0$~\cite{lancet1993proba,tanaka2009high}.

Several experimental studies have recently investigated the value of $\k$ applicable to the distribution of beneficial effects in viral or bacterial populations~\cite{Beisel:2007we,Bataillon:2014dh}. The sample sizes available in these studies are, however, insufficient to conclusively validate or invalidate the EVT hypothesis. In these experiments, the number of mutants found in the tail has indeed been so far very low, of the order of a dozen: estimating the sign of the shape parameter $\kappa$ can be attempted~\cite{Rokyta:2008kj}, but assessing the validity of the fit using quantile-quantile plots as in Figure~\ref{fig:stat} is simply impossible with such limited data. Our rich dataset thus provides the first thorough test of the applicability of EVT to the analysis of biological diversity. 

Comparable datasets are now being increasingly produced. In particular, several groups have characterized the phenotype of every single-point mutant of a protein~\cite{fowler2014deep}. Our model may be viewed as a mathematical formalization of the concept of a random library, from which single-points mutants may deviate. We note, however, that selectivities from non-random subsets of one of our libraries do follow the same model as the full library (Figure S14). In any case, significant deviations will have to be quantified against our null model.

Beyond protein libraries, the model is relevant to the screening of synthetic chemical libraries, including the combinatorial libraries of small molecules developed in the pharmaceutical industry for drug discovery~\cite{Schreiber:2009dd,Galloway:2010gs}. In this context, one previous study was performed with enough data points to possibly discriminate between different universality classes~\cite{young1997optimum}, although the authors considered only the exponential case $\k=0$.

Finally, our work raises a new question for future studies: if the selective potential of a partially randomized library is captured by few parameters, and if these parameters can vary from library to library, what controls them? More simply, what features of the framework define a universality class? For instance, how extending the variable parts to other sites changes $\k$? The patterns of amino acids forming the sequences, which we have analyzed here only to confirm the reproducibility of the experimental results and their specificity with respect to the targets and libraries, may provide valuable insights~\cite{Mora:2010jx}. 

The question may also be asked at another level: can we or natural evolution control these parameters to optimize the selective potential of a population? This question relates to the debated ``evolution of evolvability''~\cite{Wagner:1996ut,pigliucci2008evolvability}, cast here into a concrete conceptual and experimental setting. Antibodies potentially define an excellent model system to study experimentally this question, since they are subject to selection and maturation towards a diversity of targets as part of their natural function. The approach and concepts introduced in this work provide the means to address the problem with quantitative experiments.

\section*{Materials \& Methods}

{\small

\noindent {\bf Phage Display --} PVP-plates were prepared as described in~\cite{Soshee:2014ku}. The DNA target was prepared by self-assembly of a hairpin DNA, labeled with biotin at its 5' end (5'-Biotin-- AAAAGACCCCATAGCGGTCTGCGT), and was purchased from Eurogentec (Angers, France).  {\it E. coli} TG1 competent cells were purchased from Lucigen Lt.  Phage production, phage-display screens based on the pIT2 phagemid vector and  helper phage  KO7 production were performed following the standard protocol from Source BioScience (Cambridge, U.K.; http://lifesciences. sourcebio-science.com/media/143421/tomlinsonij.pdf) and our own previous work~\cite{Soshee:2014ku,Jain:2014bz} with some modifications specified in the \si.\\

\noindent {\bf Sequencing data --} Library phagemids were purified from E. coli stocks after each selection round using Midiprep kits from Macherey-Nagel (Hoerdt, France). v3 Illumina MiSeq sequencing was performed by Eurofins Genomics (Ebersberg, Germany). The MiSeq pair-ended technology was used. Frameworks were recovered on the forward read and only the reads having all the expected restriction sites and less than 4 errors on the 126 bases were kept. The CDR3 were accessible on the reverse read and only the reads having all the expected restriction sites and a average value of quality read Q$>$30 on the 12 bases defining the CDR3 were kept (see Table S2 for an estimation of sequencing errors).\\

\noindent {\bf Computational analysis --} We infer the selectivity $s_i$ of an amino acid sequence $i$ by Eq.~\eqref{eq:s_i} with $t=3$ (third round of selection). The frequencies are simply given by $f_i^{t}= n_i^{t}/\sum_j n_j^{t}$ where $n_i^{t}$ is the number of sequences $i$ present in the sample. Given sampling errors, estimated as $\Delta s_i/s_i=1/\sqrt{n_i^2}+1/\sqrt{n_i^3}$,
and given sequencing errors, estimated at $\sim$ 5\% over the 12 bases of the CDR3 (Table S2), the estimation of $s_i$ is meaningful only for sequences that are sufficiently present at each round: $n_i^{t-1}>n_0$ and $n_i^{t}>n_0$. We took $n_0=10$ and verified that the results are not sensitive to this exact value (Table S3). With $n_0=10$, relative sampling errors are, in the worst case, as high as $2/\sqrt{n_0}\sim 60\%$, but, assuming that sampling errors are uncorrelated, this uncertainty has no major incidence on the estimation of aggregated properties of the distribution of the largest $s_i$, which involves several hundreds of different $i$.\\

\noindent {\bf Extreme value statistics --} We followed the standard approach for modeling threshold excesses~\cite{coles2001introduction}. The parameters $\kappa$ and $\tau$ were estimated by maximum likelihood and the 95$\%$ confidence intervals shown in Figure~\ref{fig:stat}A were obtained under the hypothesis of normality by calculating the inverse of Fisher's information. To ensure that the data allows us to discriminate between $\k=0$ and $\k\neq 0$, a p-value was calculated by a likelihood ratio test, whose distribution was estimated both by numerical simulations. Maximum likelihood estimations are calculated on at least 50 data points. 

\section*{Author contributions}
SB, CN and OR designed research; AKS set up phage display; AKS and OR designed the libraries; DB constructed the libraries; SB performed the selections;  CN and NS supervised the experiments; SB and OR analyzed data; SB, CN and OR wrote the paper. DB, AKS and NS contributed equally to this work.

\begin{acknowledgments}
This work was supported by Agence Nationale de la Recherche (ANR-10-PDOC-004-01, to O.R.) and by AXA Research Fund (post-doctoral grant to D.B). We thank S. Girard, B. Houchmandzadeh, T. Mora, R. Ranganathan and  A. Walczak for helpful discussions.
\end{acknowledgments}

}

\newpage

\makeatletter
\newcommand{\lyxaddress}[1]{
\par {\raggedright #1
\vspace{1.4em}
\noindent\par}
}
\newenvironment{lyxlist}[1]
{\begin{list}{}
{\settowidth{\labelwidth}{#1}
 \setlength{\leftmargin}{\labelwidth}
 \addtolength{\leftmargin}{\labelsep}
 \renewcommand{\makelabel}[1]{##1\hfil}}}
{\end{list}}

\makeatletter
\makeatletter \renewcommand{\fnum@figure}
{\figurename~S\thefigure}
\makeatother

\makeatletter
\makeatletter \renewcommand{\fnum@table}
{\tablename~S\thetable}
\makeatother

\makeatletter 
\def\tagform@#1{\maketag@@@{(S\ignorespaces#1\unskip\@@italiccorr)}}
\makeatother

\appendix

\onecolumngrid

\begin{center}
{\Large SUPPORTING INFORMATION}
\end{center}

\section*{\large Supplementary experimental methods}

\subsection*{Library construction} 

The library-specific parts of the frameworks, upstream of the variable CDR3 (Figure 1) are shown in Figure~S\ref{fig:seq_fram}. 23 of these frameworks were designed based on the amino acid sequences of 23 natural $V_H$ segments, with minor modifications to accommodate common restriction sites at the two ends of the CDR2 and CDR3. Out of these 23 frameworks, 20 were chosen to have minimal sequence similarities, and 3 are from a same human $V_H$ segment: one is the germline (na\"ive) form, one results from limited maturation (85\% sequence similarity to the germline) and the other from extensive maturation (broadly neutralizing antibody against HIV~\cite{Klein:2013iz} with only 65\% sequence similarity to the germline). A 24th framework was made exclusively of glycines to serve as a control. Downstream of the CDR3, the fixed part has amino acid sequence FDYWGQGTLVTVSSG in all libraries. The nucleotide sequences were optimized for {\it E. coli} codon usage and are provided as supplementary file.\\

The 24 frameworks were obtained from Genewiz (South Plainfield, NJ) as synthetic genes with restriction sites flanking the CDRs to allow for the introduction of arbitrary sequences at the CDRs. In particular, the CDR3 region is flanked by BssHI and XhoI sites. These synthetic genes were cloned into a modified version of pIT2 phagemid (standard phage display vector) lacking $V_L$~\cite{Soshee:2014ku}. To randomize the CDR3 region, a degenerate oligonucleotide containing 12 random nucleotides (from Eurogentec, Angers, France) flanked by BssHI and XhoI sites was PCR-amplified, digested, and ligated into gel purified pIT2 phagemids harboring each of the 24 frameworks. Ligation products were purified and electroporated into TG1 {\it E. coli} (from Lucigen, Middleton, WI) at efficiencies exceeding $10^{7}$ transformants (to ensure a $>$100-fold coverage of the $10^{5}$ diversity), while keeping 100-fold lower efficiency in control electroporations of ligation product without insert (to minimize the occurence of empty vectors in libraries, below 1\%).

\subsection*{Phage display screening}

All chemicals were purchased from Sigma-Aldrich (St Louis, MO) unless otherwise specified. Deionized water of resistivity 16 M$\Omega$.cm was produced with an ion exchange resin (Aquadem(R) system, Veolia, Lyon, France). 2xTY medium was prepared by dissolving 16 g tryptone, 10 g yeast extract, 5 g NaCl (tryptone and yeast extract from USBIO distributed by Euromedex, Strasbourg, France) in 1 L of deionized H2O and autoclaving for 15 min at 120 $^{\circ}$C.\\

The DNA target (PAGE purified, lyophilized) was resuspended with deionized water at 400 $\mu$M. 20 mg of magnetic beads coated with streptavidin (Dynabeads(R) M-280 Streptavidin from Invitrogen Life Technologies SAS, Saint Aubin, France) were prepared according to the manufacturer's protocol. 10 $\mu$L of DNA stock solution were mixed with 20 mg of washed Dynabeads(R) and incubated for 10 minutes at room temperature using gentle rotation. The biotinylated hairpin DNA coated beads were separated with a strong magnet for 2-3 minutes and washed 2-3 times with a buffer containing 5 mM Tris-HCl (pH 7.5), 0.5 mM EDTA and 1M NaCl.\\

Phage production is the same as described before except that the infected TG1 culture was grown for 7 hours (instead of overnight) at 30$^{\circ}$C in 2xTY + 100 $\mu$g/mL ampicillin + 50 $\mu$g/mL kanamycin.\\

During phage display experiments, the supernatant containing our library (around $10^{11}$ phages) in 2xTY, was adjusted to 10 mM NaPO4 pH=7.4. Phages were first incubated against either naked magnetic beads or non-treated polystyrene 3 cm diameter Nunc Petri dish (Thermo Fisher Scientific, Waltham, MA) for negative selection. For DNA target selection, DNA LoBind tubes (Eppendorf AG, Hamburg, Germany) were used. Phages were incubated during 1 h without agitation and 30 min on a rocker at room temperature. The remaining phages were then incubated with either hairpin DNA or PVP targets. In the case of hairpin DNA, 50 $\mu$L of beads were incubated with an excess of DNA targets (around $10^{14}$), washed according to the manufacturer's protocol,  yielding on the order of $10^{13}$ immobilized DNA targets, at a 100-fold excess over available phages ($10^{11}$). Antibody selection was then performed against either DNA-coated beads or a PVP-functionalzed Petri dish for 90 mn on a rocker. 10 washing steps with 1xPBS + 0.1$\%$Tween 20 were performed. Next, selected phages were eluted using 1 mL of fresh solution of 100 mM triethylamine for 20 min and neutralized with 500 $\mu$L of Tris/HCl buffer (1 M, pH 7.4). Eluted phages were rescued by infection of an excess of exponentially growing TG1 {\it E. coli} cells (14 mL of a 2xTY culture at O.D. 600 nm = 0.6) for titration and phage preparation for subsequent rounds of selection. Infected TG1 were then plated on 2xTY + ampicillin plates for overnight amplification at 37$^{\circ}$C. Glycerol stocks were stored at $-80^{\circ}$C.

\subsection*{Amplification biases}

Each round of selection is followed by a round of amplification consisting in infecting the bacteria with the selected phages. Sequence-specific differences in amplification may arise from differences of growth rate of the bacteria carrying different phagemids, or differences in infectivity or display ability of the phages. We measured by sequencing both the differences between frameworks when considering a mixture of the 24 libraries (Figure S\ref{fig:amp1}) and between CDR3 when considering a library of given framework (Figure S\ref{fig:amp2}).\\

Between frameworks, only the S1 and CH1 libraries, show significant enrichment upon amplification alone. Each of these two libraries dominates over the others in one experiment of selection with a mixture of libraries but, when the mixture of all 24 libraries is selecting against the PVP target, they are dominated by another library, the HG library, which does not show any enrichment upon amplification. This observation, together with the strong correlation between frequencies before and after amplification (Figure S\ref{fig:amp1}B), are evidence that differences in library amplification are not responsible for the observed hierarchy between frameworks (Figure 2).\\

Within each library, a clear enrichment for the glutamine amino acid is observed, irrespectively of the framework (Figure S\ref{fig:amp2}). This bias has a simple interpretation: the {\it supE} strain of {\it E. Coli} that we use for phage display is a partial amber stop codon suppressor. In this strain, the amber codon codes about one third of the times for a glutamine and acts as a stop codon the two other thirds. The reduced production of antibodies due to the presence of an amber codon thus confers a growth advantage to the bacteria (antibody expression is costly for {\it E. coli}). Consistently with this interpretation, we verify that all the glutamines present in the data are associated with the amber codon. The results presented in the paper exclude sequences with an amber codon but, in most experiments with selection, glutamine does not appear in the selected consensus sequence and considering the amber code as coding for an amino acid or for a stop codon has no incidence on the conclusions. Apart from glutamine amplification, no other significative pattern of amplification is visible or may plausibly explain the results of the experiments with selection.

\vspace{.5cm}

\section*{\large Supplementary properties of extreme value distributions}

\subsection*{Relations between parameters}

The fit to an extreme value distribution with parameters ($\k_0,\tau_0$) applies for selectivities above a threshold $s^*_0$. Fitting the data above a larger threshold $s^*_1>s^*_0$ must lead to the same shape parameter $\k_1=\k_0$ (simply denoted $\k$) but to a different scaling parameter $\tau_1$ given by~\cite{coles2001introduction}
\beq\label{eq:tau}
\tau_1=\tau_0+\k(s_1^*-s_0^*).
\eeq
These are the relations verified in Figure 4A.\\

Another independent parameter, which depends on the bulk of the distribution, is the fraction $\phi_0$ of the data above the threshold $s^*_0$, which obviously depends on the value of the threshold ($\phi_1<\phi_0$ when $s^*_1>s^*_0$).\\

In total, four parameters are thus relevant: $s^*$, $\phi$, $\k$ and $\tau$.

\subsection*{Spacings between extremes}

We show here that if $s_1>s_2>\dots$ are drawn at random with a probability density $f_{\kappa,\tau,s^*}(s)$ given by Eq.~(3) then their spacings defined by $\Delta_r=s_r-s_{r+1}$ scale as $\Delta_r/\Delta_1\sim r^{-(\k+1)}$, where $\k$ is the shape parameter. This follows from a more general result:
\beq\label{eq:spacing}
\Delta_r\sim\tau N^\k r^{-(\k+1)},
\eeq
where $\tau$ is the scaling parameter and $N$ the total number of samples.\\

The proof can be given in terms of the rescaled variable $x=(s-s^*)/\tau$ whose probability density $f_\k(x)$ is defined in Eq. (4), since $\Delta_r=\tau (x_r-x_{r+1})$. As indicated by Eq. (5), the rank $r$ and the value $x$ are related for large $N$ by $r(x)/N\sim\int_x^\infty f_\k(u) {\rm d}u=(1+\k x)^{-1/\k}$. Inverting this relation gives $x_r=q(r/N)$ with $q(z)=(1-z^{-\k})/\k$. In the limit of large $N$ where the formalism applies, we have therefore $x_r-x_{r+1}\simeq -(1/N)q'(r/N)$ with the derivative of $q(z)$ given by $q'(z)=-z^{-(\k+1)}$. All together, this gives $x_r-x_{r+1}\simeq N^\k r^{-(\k+1)}$ and thus $\Delta_r\sim\tau N^\k r^{-(\k+1)}$.

\subsection*{Scaling of maxima with library sizes}

We show here that if the maximum of $N$ random variables drawn with a probability density given by Eq.~(3) is $s_1$, then adding more elements to produce a library $m>1$ times larger leads to a maximal value $s'_1\geq s_1$ satisfying
\beq\label{eq:scaling}
\E[s_1'-s_1]=\tau N^\k m^\k \Li_{\k+1}\left(1-\frac{1}{m}\right),
\eeq
where $\Li_k(z)=\sum_{n=1}^\infty z^n/n^k$ defines the so-called polylogarithmic function. $\E[s_1'-s_1]$ is thus an increasing function of $\kappa$ as illustrated in Figure~S\ref{fig:scaling}, where Eq.~\eqref{eq:scaling} is also compared to numerical simulations. The relevance of this formula rests on the assumption that sub-libraries of a library with shape parameter $\k$ are characterized by the same $\k$, which finds support in the data (Figure~S\ref{fig:sub}). In practice, the extreme value distribution applies only to the fraction $\phi N$ of the data above the threshold $s^*$. When expressed relative to the expected spacing between the top two values in the initial population of $N$ variables, $\Delta_1=\tau N^\k$, Eq.~\eqref{eq:scaling} is however independent of $N$ and $\tau$: $\E[s_1'-s_1]/\Delta_1= m^\k \Li_{\k+1}\left(1-\frac{1}{m}\right)$.\\

To derive this formula, we consider an initial population of size $mN$ whose maximum is $s_1'$ and define a subpopulation of (approximate) size $N$ by retaining with probability $1/m$ each of its elements. The maximum $s_1$ of this subpopulation has thus rank $n$ in the initial population with probability $p_n=(1-1/m)^{n-1}1/m$ -- the probability that none of $n-1$ top values are retained but that the $n$-th is. Following Eq.~\eqref{eq:spacing}, the distance between $s_1=s'_n$ and $s'_1$ is estimated as $\delta'_n=s_1'-s_n'=\sum_{r=1}^{n-1}\Delta_r'=\tau (mN)^\k\sum_{r=1}^{n-1}r^{-(\k+1)}$. This leads to
\beq
\E[s_1'-s_1]=\sum_{n=1}^\infty p_n\delta_n'=\frac{\tau (mN)^\k}{m}\sum_{n=1}^\infty\sum_{r=1}^{n-1} \left(1-\frac{1}{m}\right)^{n-1}\frac{1}{r^{\k+1}}
=\frac{\tau (mN)^\k}{m}\sum_{r=1}^{\infty}\sum_{n=r+1}^\infty \left(1-\frac{1}{m}\right)^{n-1}\frac{1}{r^{\k+1}}
\eeq
and, after summing the geometric series $\sum_{n=r+1}^\infty \left(1-1/m\right)^{n-1}=(1-1/m)^r m$, to
\beq
\E[s_1'-s_1]=\tau (mN)^\k\sum_{n=1}^\infty  \left(1-\frac{1}{m}\right)^{n-1}\frac{1}{r^{\k+1}},
\eeq
which is equivalent to Eq.~\eqref{eq:scaling}.

\vspace{1cm}

\section*{\large Supplementary tables}

\begin{table*}[h]
\centering
{\small
\begin{tabular}{l|l|l|l|l|}
\cline{2-5}
                                     & $\kappa$         & $\tau$                                    & $s^{*}$               & $\phi$           \\ \hline
\multicolumn{1}{|l|}{S1/PVP}  & $0.44 \pm 0.22$  & $1.6\times 10^{-4} \pm 10^{-5}$           & $0.37 \times 10^{-3}$ & $\sim 6\times10^{-4}$   \\ \hline
\multicolumn{1}{|l|}{F3/PVP}  & $0.07 \pm 0.21$  & $3.1\times 10^{-4} \pm 8\times 10^{-5}$   & $1.2 \times 10^{-3}$  & $\sim 6\times 10^{-4}$  \\ \hline
\multicolumn{1}{|l|}{HG/DNA}  & $0.26 \pm 0.21$  & $5.7 \times10^{-3} \pm 1.5\times 10^{-3}$ & $0.7 \times 10^{-3}$  & $\sim 6\times 10^{-4}$  \\ \hline
\multicolumn{1}{|l|}{CH1/DNA} & $-0.62 \pm 0.25$ & $2.5 \times10^{-2} \pm 8\times 10^{-3}$   & $2.5 \times 10^{-3}$  & $\sim 3.6\times10^{-4}$ \\ \hline
\end{tabular}
}
\caption{Parameters $\k$, $\tau$, $s^*$ and $\phi$ describing the four experiments presented in Figure 3. Note that $\tau$ and $\phi$ depend on $s^*$, which may be chosen within a finite interval of values. However, the values of $\tau(s^*)$ and $\phi(s^*)$ at $s^*=s^*_0$ determine their values at $s^*=s^*_1$ as indicated in Eq.~\eqref{eq:tau} for $\tau$. \label{tab:recap}}
\end{table*}

\begin{table}[h]
{\small
\centering
\begin{tabular}{l|l|}
\cline{2-2}
                                                         & \begin{tabular}[c]{@{}l@{}}Fraction of sequences with\\ $>$1 sequencing error / 12 bases\end{tabular} \\ \hline
\multicolumn{1}{|l|}{Mix24}                              & 0.043                                                                                                                 \\ \hline
\multicolumn{1}{|l|}{Mix24 amplified}                    & 0.029                                                                                                                 \\ \hline
\multicolumn{1}{|l|}{Mix24/PVP round 1}           & 0.025                                                                                                                 \\ \hline
\multicolumn{1}{|l|}{Mix24/PVP round 2}           & 0.051                                                                                                                 \\ \hline
\multicolumn{1}{|l|}{Mix24/PVP round 3}           & 0.107                                                                                                                 \\ \hline
\multicolumn{1}{|l|}{Mix24/DNA round 1}           & 0.032                                                                                                                 \\ \hline
\multicolumn{1}{|l|}{Mix24/DNA round 2}           & 0.065                                                                                                                 \\ \hline
\multicolumn{1}{|l|}{Mix24/DNA round 3}           & 0.029                                                                                                                 \\ \hline
\multicolumn{1}{|l|}{Duplicate Mix24/DNA round 3} & 0.024                                                                                                                 \\ \hline
\multicolumn{1}{|l|}{Mix21/PVP round 2}       & $4\times 10^{-3}$                                                                                                                 \\ \hline
\multicolumn{1}{|l|}{Mix21/PVP round 3}       & $4\times 10^{-3}$                                                                                                                     \\ \hline
\multicolumn{1}{|l|}{Mix21/DNA round 1}           & 0.027                                                                                                                 \\ \hline
\multicolumn{1}{|l|}{Mix21/DNA round 2}           & 0.046                                                                                                                 \\ \hline
\multicolumn{1}{|l|}{Mix21/DNA round 3}           & 0.106                                                                                                                 \\ \hline
\multicolumn{1}{|l|}{F3/PVP round 1}           & 0.034                                                                                                                 \\ \hline
\multicolumn{1}{|l|}{F3/PVP round 2}           & 0.029                                                                                                                 \\ \hline
\multicolumn{1}{|l|}{F3/PVP round 3}           & 0.04                                                                                                                  \\ \hline
\multicolumn{1}{|l|}{F3/DNA round 1}           & 0.027                                                                                                                 \\ \hline
\multicolumn{1}{|l|}{F3/DNA round 2}           & 0.048                                                                                                                 \\ \hline
\multicolumn{1}{|l|}{F3/DNA round 3}           & 0.085                                                                                                                 \\ \hline
\end{tabular}
}
\caption{Estimation of sequencing errors -- Fraction of the sequences with at least one error in the 12 bases immediately downstream of the 12 based of the CDR3 (errors estimated given the known sequence of the fixed part of the framework).\label{tab:err}}
\end{table}

\begin{table}[h]
\centering
{\small
\begin{tabular}{|l|l|}
\hline
S1/PVP in Mix24 $n_{0}^2=n_0^3=10$                        & $\kappa=0.34 \pm 0.22$ \\ \hline
S1/PVP in Mix24 $n_{0}^{2}=10$ and $n_{0}^{3}=25$ & $\kappa=0.42 \pm 0.25$ \\ \hline
S1/PVP in Mix21 $n_{0}^2=n_0^3=100$                       & $\kappa=0.56 \pm 0.18$ \\ \hline
S1/PVP in Mix21 sampled $n_{0}^2=n_0^3=10$                & $\kappa=0.48 \pm 0.16$ \\ \hline
S1/PVP in Mix21 sampled  $n_{0}^{2}=10$ and $n_{0}^{3}=50$                & $\kappa=0.67 \pm 0.37$ \\ \hline
\end{tabular}
}
\caption{Robustness of the EVT analysis -- The analysis presented in the main text retains only sequences present in sufficient number in the samples of the populations that are sequenced at the second and third rounds -- namely $n_i^2>n_0^2=10$ and $n_i^3>n_0^3=10$. This table shows that varying the values of the thresholds $n_0^2$ and $n_0^3$ has little incidence on the value of the shape parameter $\kappa$ inferred by EVT analysis. The sample of the S1 library against PVP in the mixture of 21 libraries (Mix21, see Figure 2) contained $10^6$ sequences while the sample in the mixture of 24 libraries (Mix24) contained only $10^5$; the two last rows of the table shows that further sampling at 1/10 the former to reach samples of comparable sizes have no incidence on the results.\label{tab:robust}}
\end{table}

\clearpage

\section*{\large Supplementary figures}

\begin{figure}[h]
\centerline{\includegraphics[width=.6\linewidth]{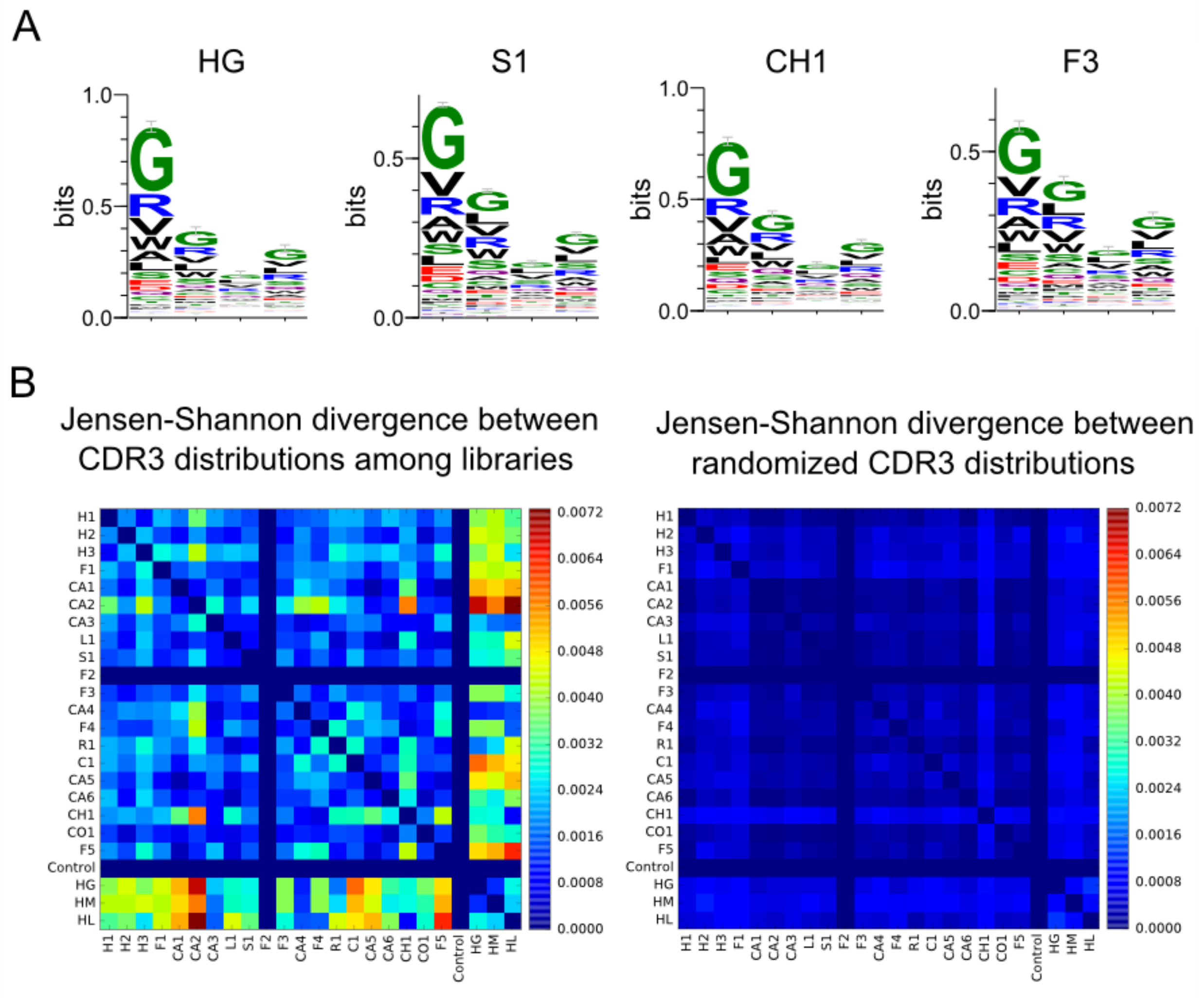}}
\caption{Diversity of the libraries -- The different libraries are intended to harbor the same distribution of amino acids at the 4 varied positions. We measured these distributions by sequencing samples from the initial libraries. {\bf A.} Sequence logos showing the entropies of the various amino acids at the four positions: the distribution is non-uniform but similar from one library to the next. {\bf B.}~More quantitatively, the distance between distributions is estimated using the Jensen-Shannon divergence: if $q_a^\ell$ is the frequency of amino acid $a$ in the CDR3 of library $\ell$, the Jensen-Shannon divergence between libraries $k$ and $\ell$ is defined as $\sum_a q_a^k\ln (q_a^k/q_a^\ell)+\sum_a q_a^\ell\ln (q_a^\ell/q_a^k)$. This divergence is found to be 5 to 10 times larger than expected from sampling noise. This represents the experimental precision at which we were able to introduce the same diversity in each library. These differences of frequencies between initial libraries are, however, much smaller than the differences of frequencies before and after a round of selection within a same library.\label{fig:div}}
\end{figure}

\begin{figure}[h]
\centerline{\includegraphics[width=.45\linewidth]{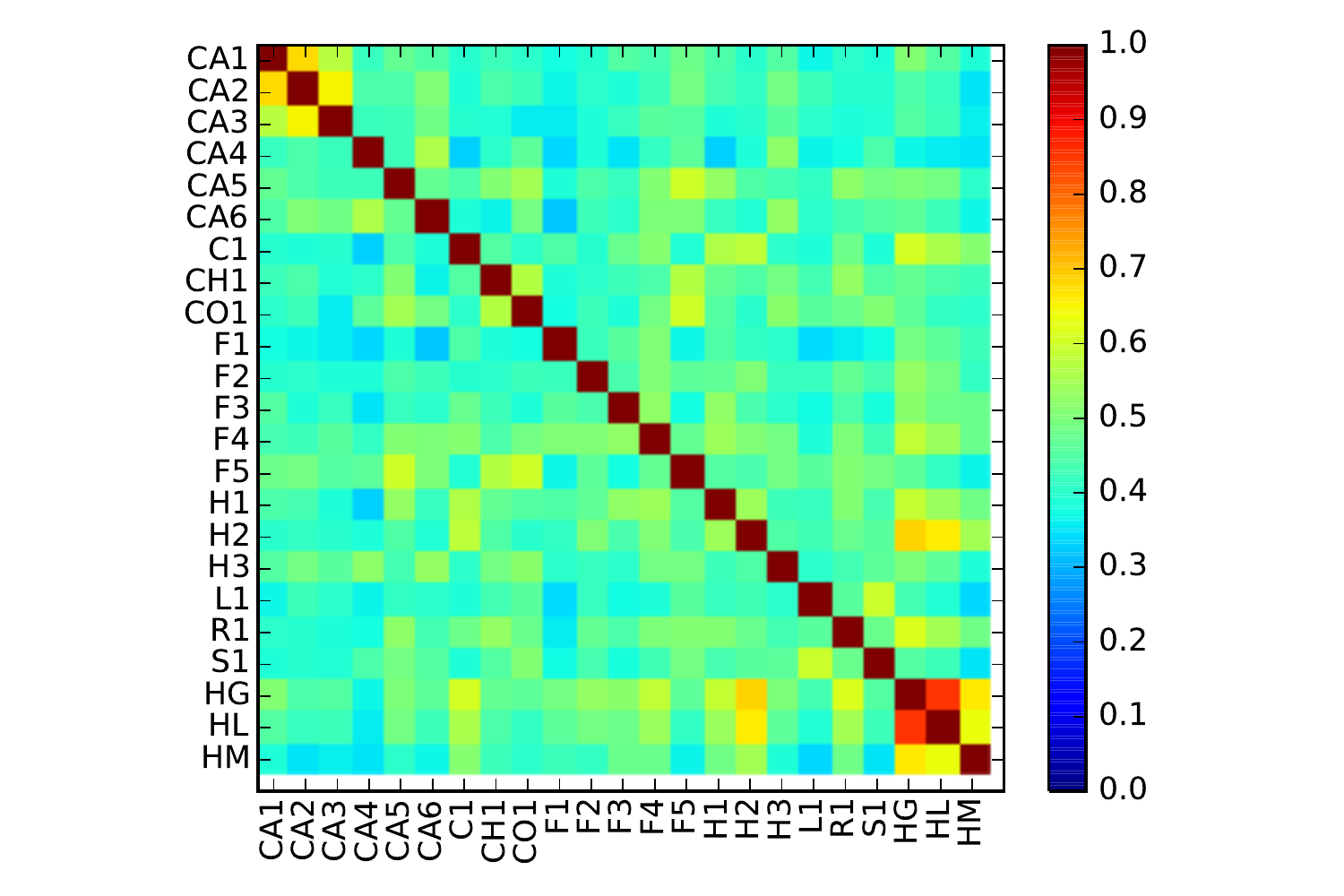}}
\caption{Sequence similarities between frameworks -- Similarity between two frameworks is measured as the fraction of common amino acids in an alignment of their two sequences. Only the library-specific part of the frameworks (Figure 1) defined in Figure~S\ref{fig:seq_fram} is considered here. In most cases, the sequence similarity is in the range 30-60\%.\label{fig:similarity}}
\end{figure}

\begin{figure}[h]
\centerline{\includegraphics[width=.4\linewidth]{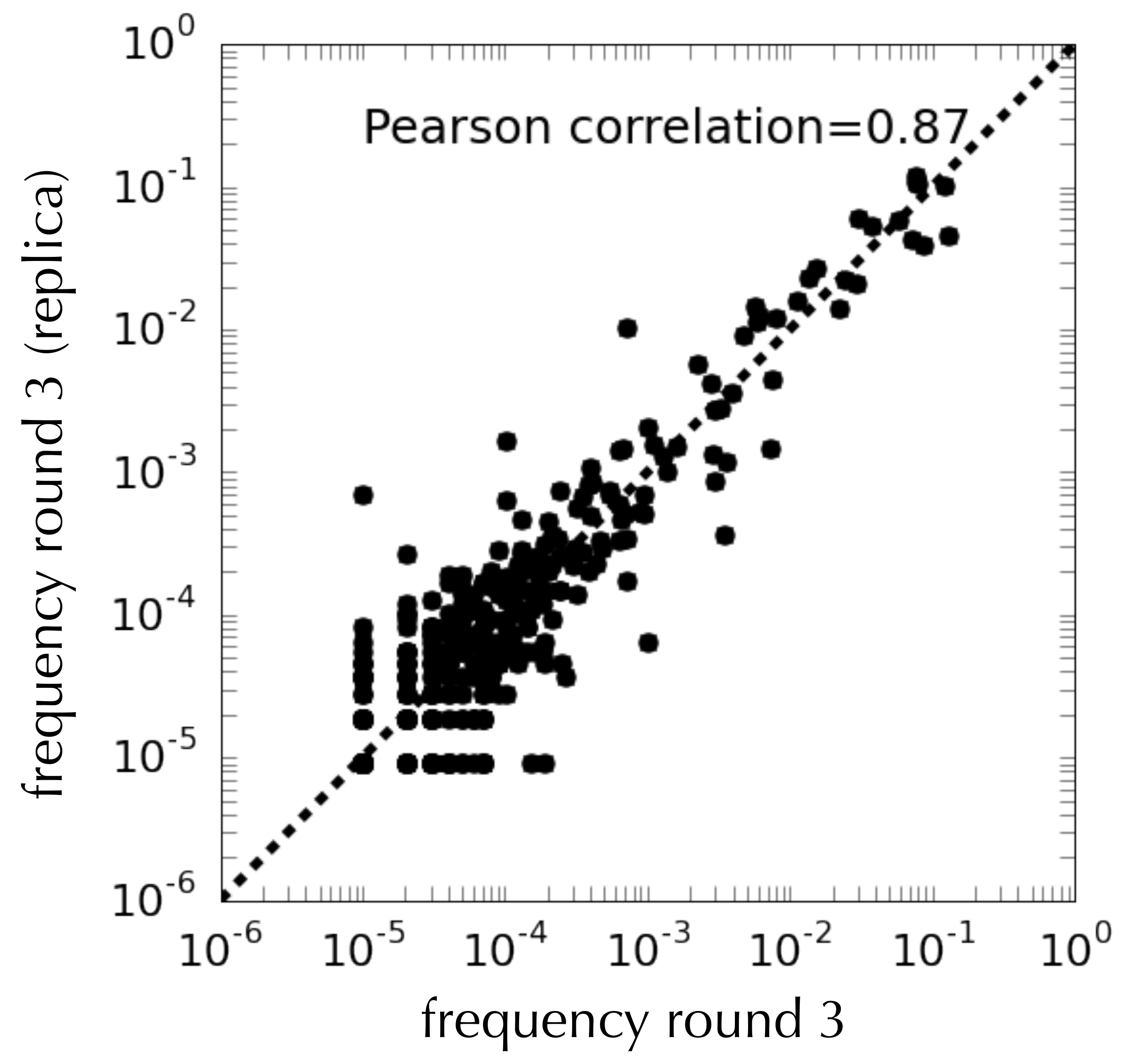}}
\caption{Reproducibility -- To assay the reproducibility of the experiments, two independent selections of the mixture of 24 libraries were performed against the DNA target and the frequencies of the sequences were compared at the third round: the high correlation between the two results indicates high reproducibility.\label{fig:repro}}
\end{figure}

\begin{figure}[h]
\centerline{\includegraphics[width=.55\linewidth]{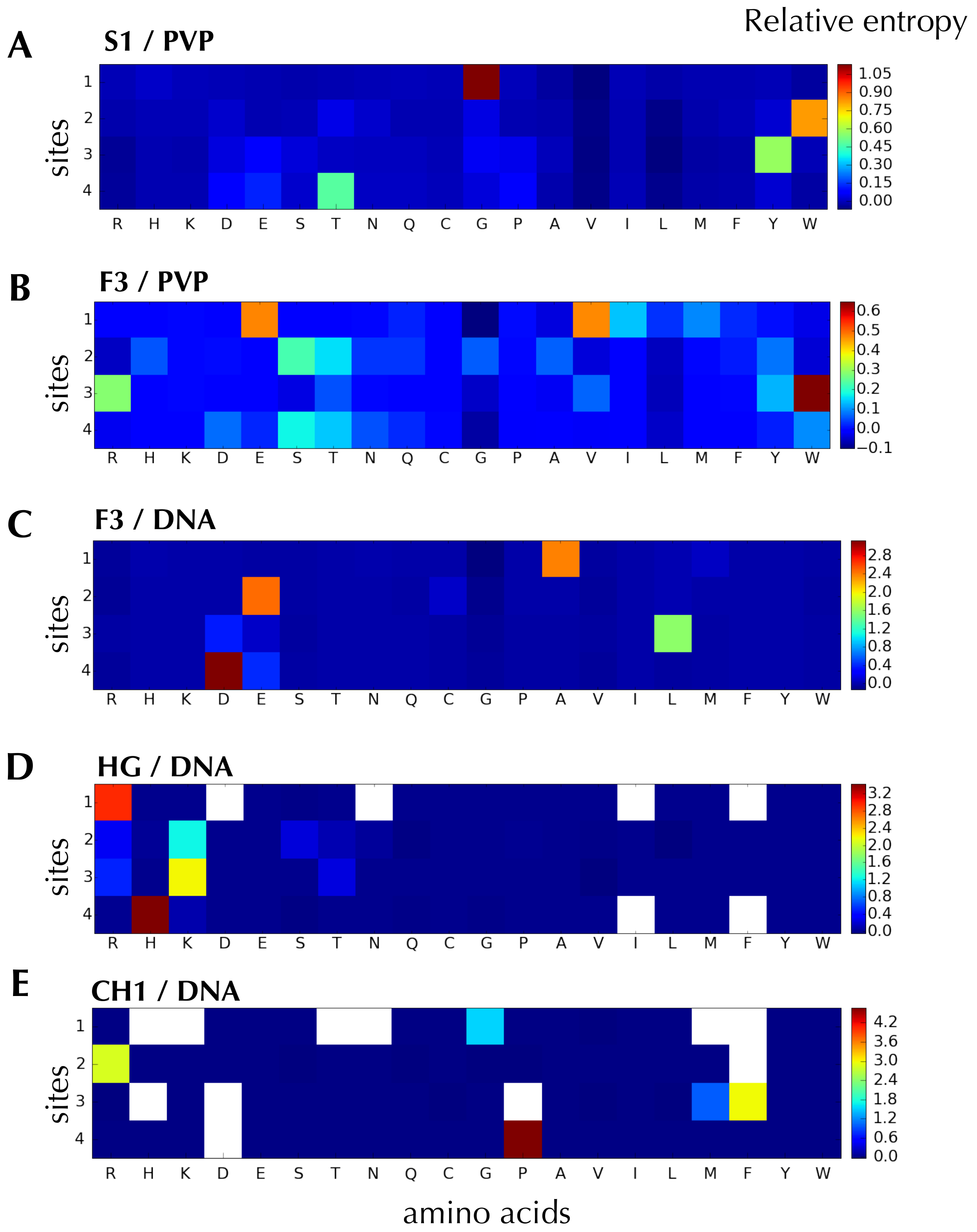}}
\caption{Library and target specificities -- Relative entropies of the different amino acids at the third round of different experiments; the relative entropy is calculated per site as $f_i^a\ln(f_i^a/q_i^a)$ where $f_i^a$ is the frequency of amino acid $a$ at position $i$ in the third round and $q_i^a$ in the initial library (round 0). \textbf{A} and \textbf{B} show that the consensus sequence is framework dependent. \textbf{B} and \textbf{C} show that it is target specific. Finally, \textbf{C}, \textbf{D} and \textbf{E} provide further evidence of framework dependency. (White squares indicate amino acid not represented in the population).\label{fig:aa1}}
\end{figure}

\begin{figure}[h]
\centerline{\includegraphics[width=.9\linewidth]{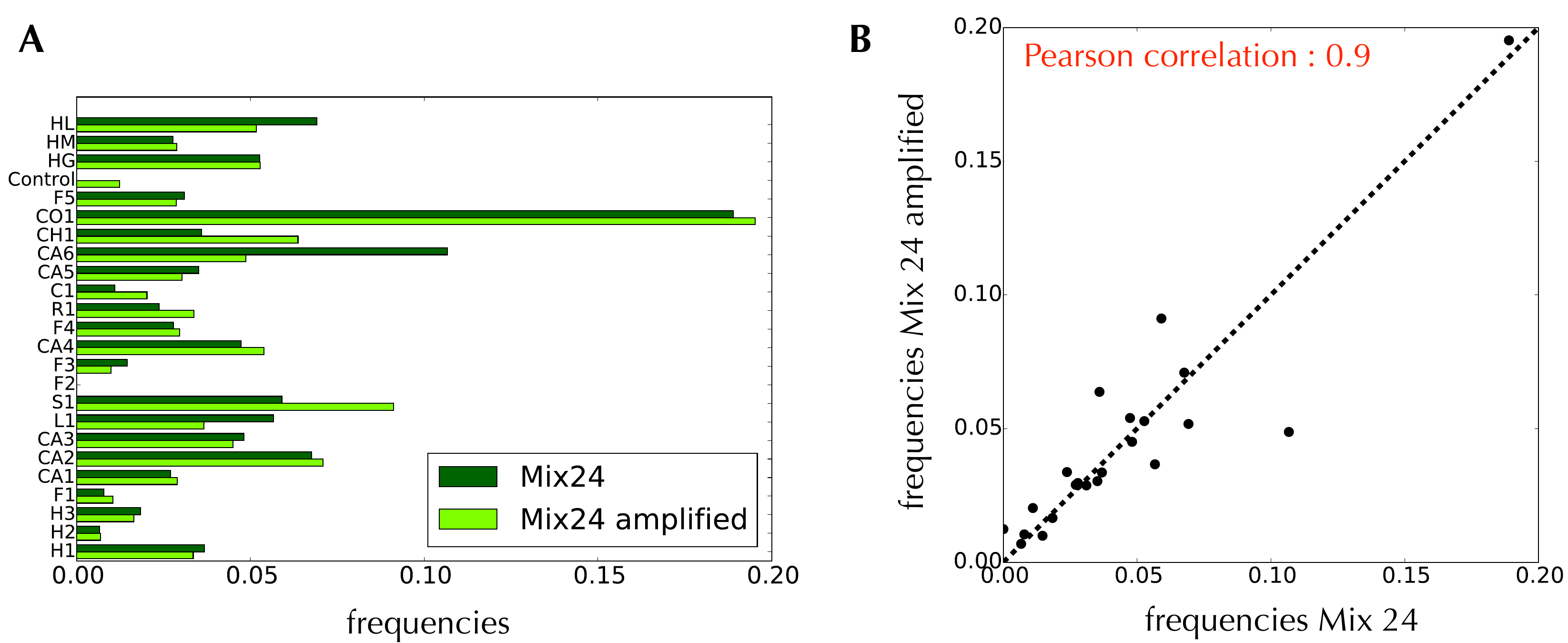}}
\caption{Biases in amplifications -- Comparison of the composition of the mixture of 24 libraries before and after amplification in absence of selection. \textbf{A.} Differences of frequencies, showing that only the S1 and CH1 libraries are enriched. \textbf{B.} Correlations between the frequencies (same data).\label{fig:amp1}}
\end{figure}

\begin{figure}[h]
\centerline{\includegraphics[width=.9\linewidth]{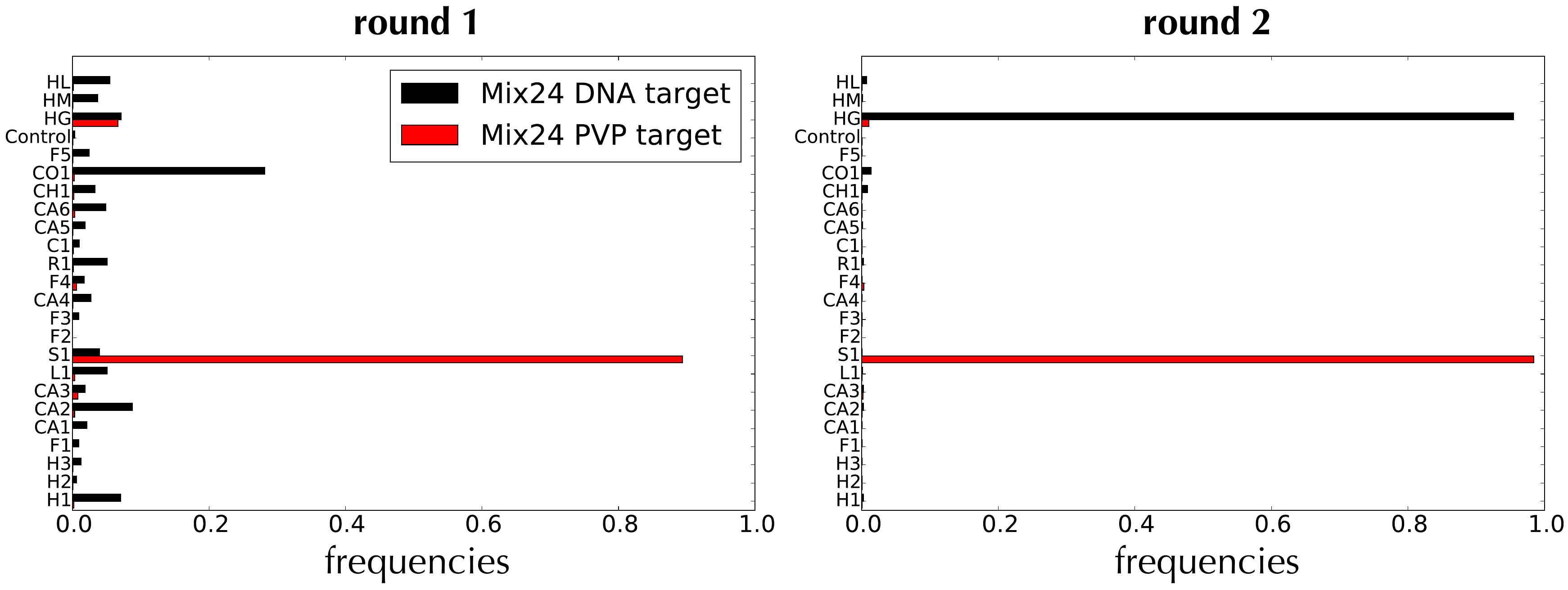}}
\caption{Target-dependent hierarchy -- Figure 2 shows that a mixture of 24 libraries selected against the DNA target is dominated by the HG framework while a mixture of 21 libraries that excludes the HG library is dominated by the CH1 framework. As shown in this figure, when the same mixture of 24
 libraries is selected against the PVP target, a different framework, the S1 framework, dominates (consistently, it also dominates when screening the mixture of 21 libraries, which includes S1).\label{fig:meta}}
\end{figure}

\begin{figure}[h]
\centerline{\includegraphics[width=.5\linewidth]{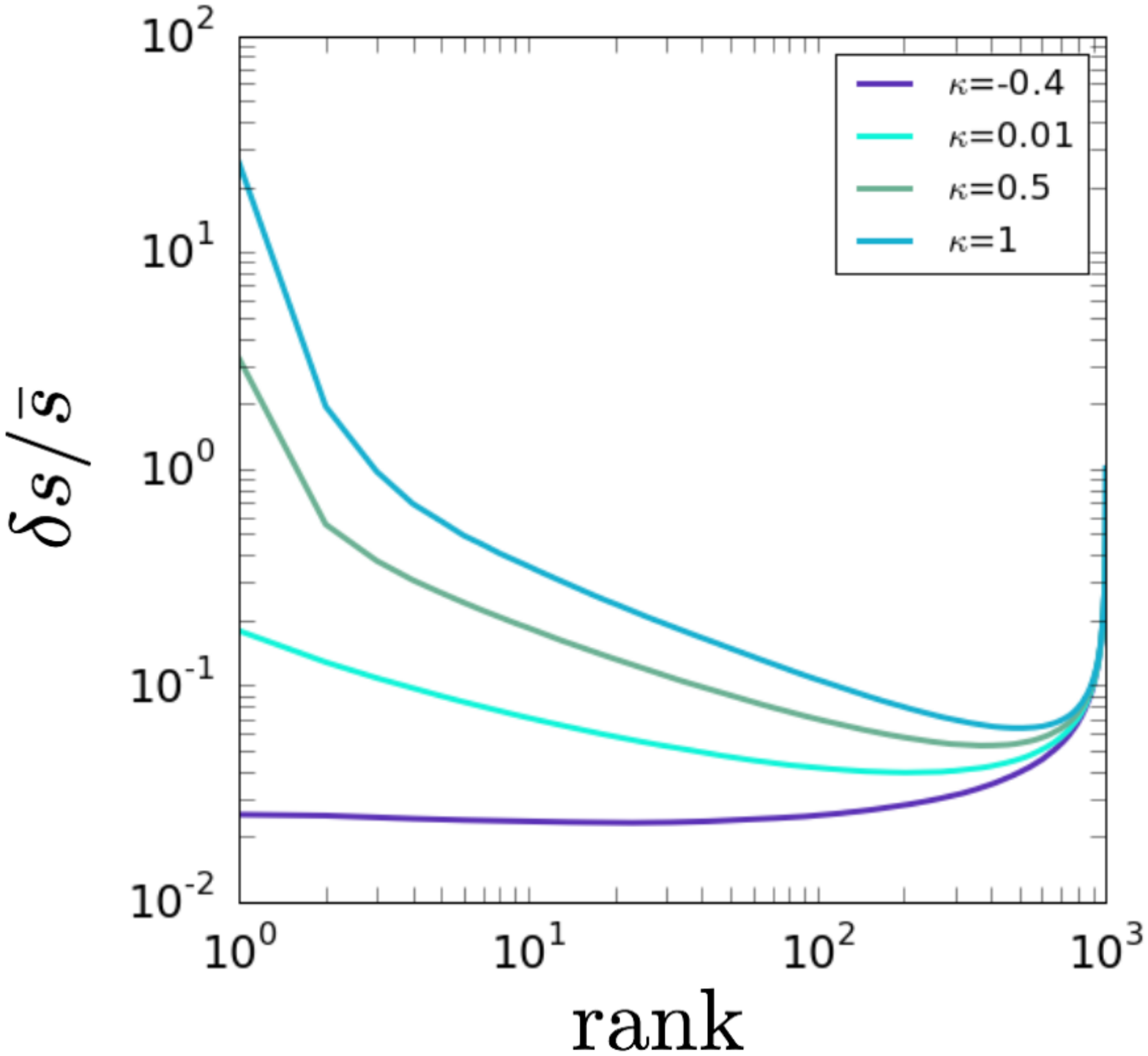}}
\caption{Variations in the values of extreme selectivities -- When sampling $N$ random variables from the extreme probability density $f_\k(x)$ given by Eq.~(4), the value $s_r$ of the variable of rank $r$ is distributed with a mean $\bar s_r$ and standard deviation $\delta s_r$. The ratio $\delta s_r/\bar s_r$ is largest for the very top sequences, as shown here based on numerical simulations. This observation is consistent with deviations of the data from a power law observed for the very top selectivities even when $\k>0$ and when the overall fit with an extreme value distribution is good (Figures 4A-B).\label{fig:var}}
\end{figure}

\begin{figure}[h]
\centerline{\includegraphics[width=.7\linewidth]{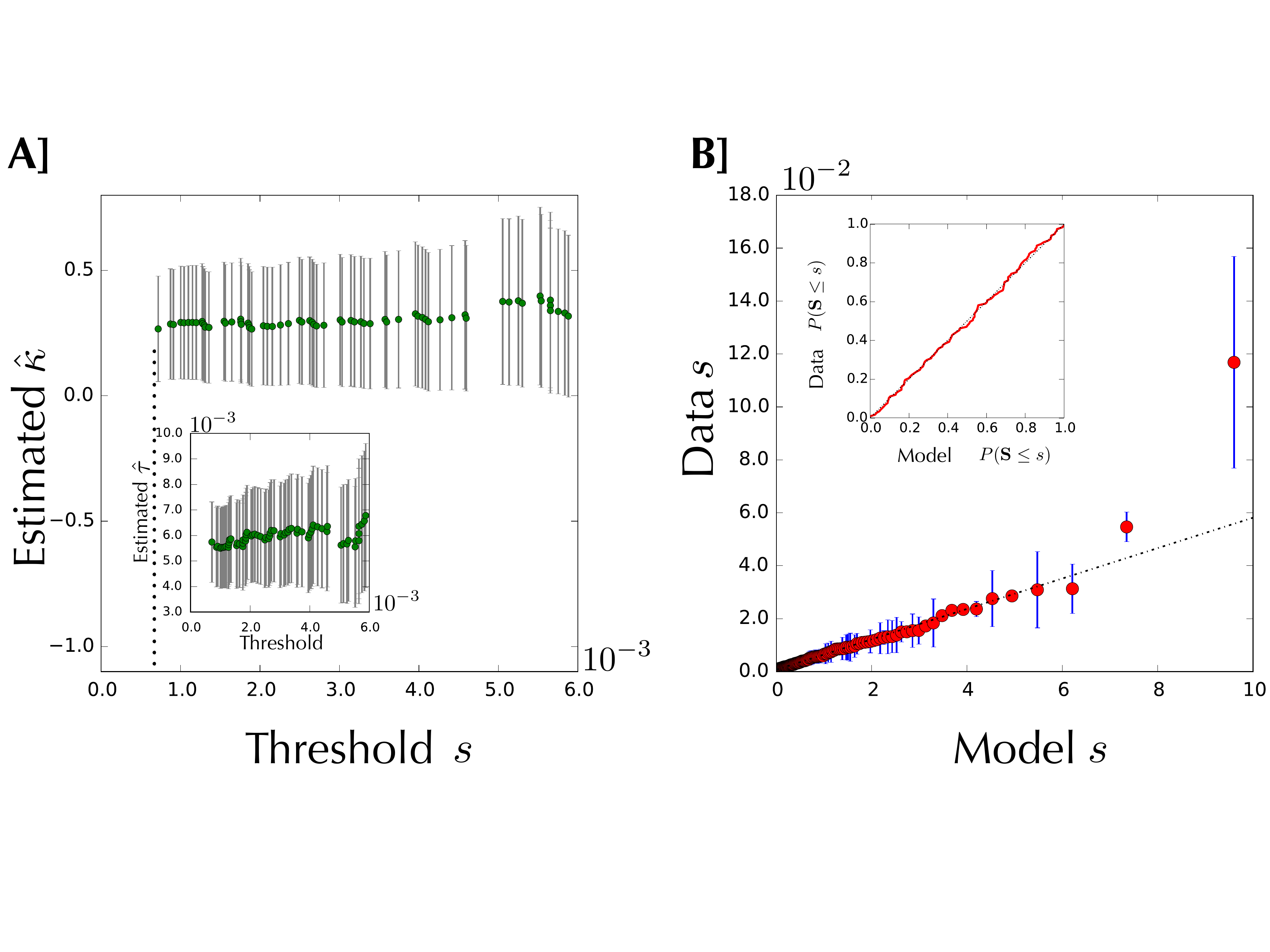}}
\caption{EVT analysis for the selection of the HG library against the DNA target (data shown in Figure 3B). A fit of the general model gives $\kappa=0.26\pm 0.21$, $\tau=5.7\times10^{-3}\pm 1.5\times10^{-3}$ while a fit of the exponential model ($\k=0$) gives $\tau_0=8\times10^{-3}\pm 1.4\times10^{-3}$; the exponential model is excluded with a p-value $1.4\times 10^{-3}$, in favor of $\k>0$. Note that the threshold $s^*=10^{-3}$ above which the fit is stable and good is much below the value of the selectivity above which a power law is observed in Figure 3B, of the order of $s= 10^{-2}$. \label{fig:germline}}
\end{figure}

\begin{figure}[h]
\centerline{\includegraphics[width=.7\linewidth]{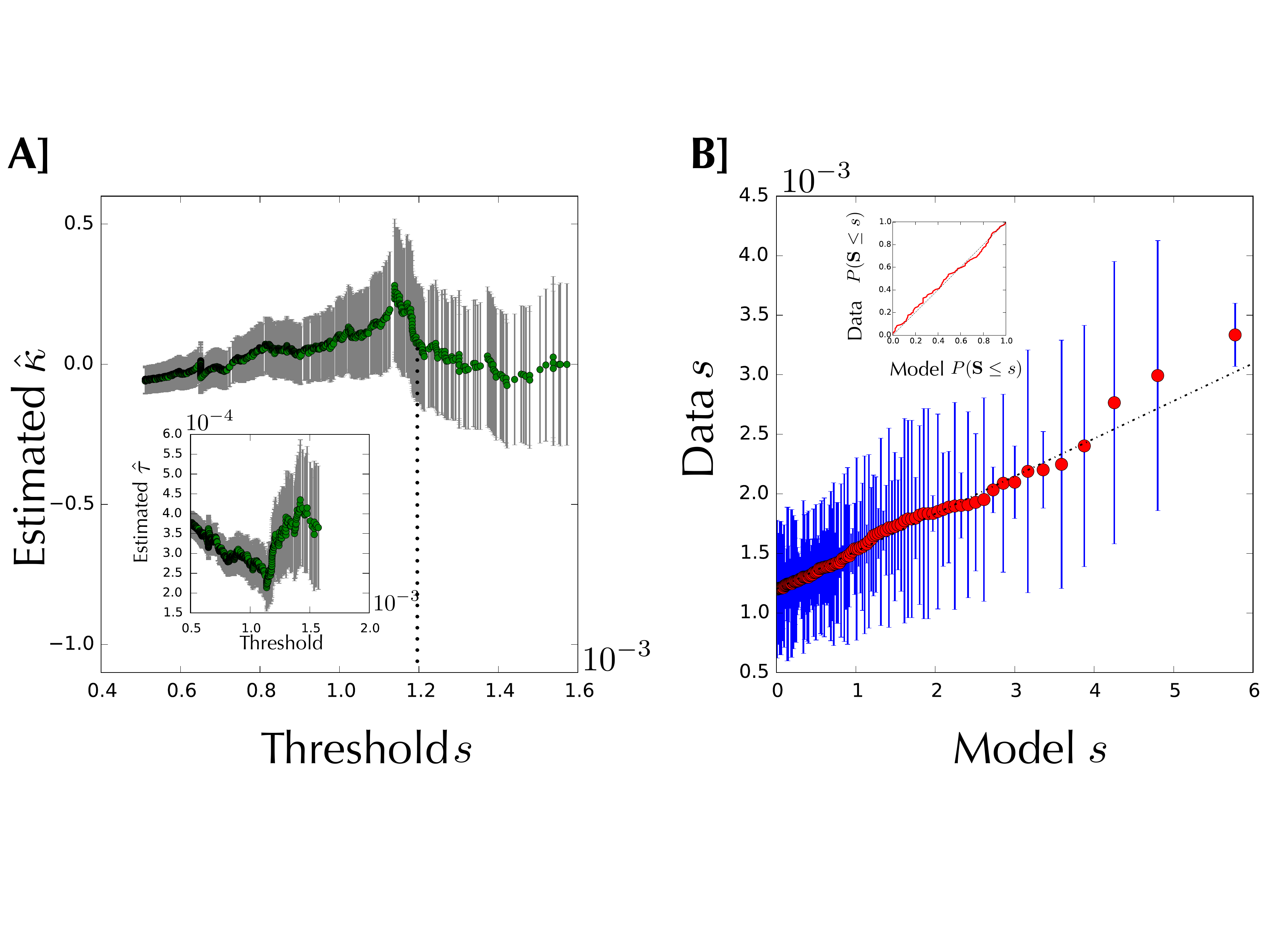}}
\caption{EVT analysis for the selection of the F3 library against the PVP target (data shown in Figure 3D). A fit of the general model gives $\kappa=0.07\pm 0.21$, $\tau=3.1\times10^{-4}\pm 8\times10^{-5}$ while a fit of the exponential model ($\k=0$) gives $\tau_0=3.4\times10^{-4}\pm 6\times10^{-5}$; the exponential model is excluded with a p-value $0.75$, which is non significant. This data is therefore consistent with an exponential model $\k=0$.\label{fig:frog3}}
\end{figure}

\begin{figure}[h]
\centerline{\includegraphics[width=.7\linewidth]{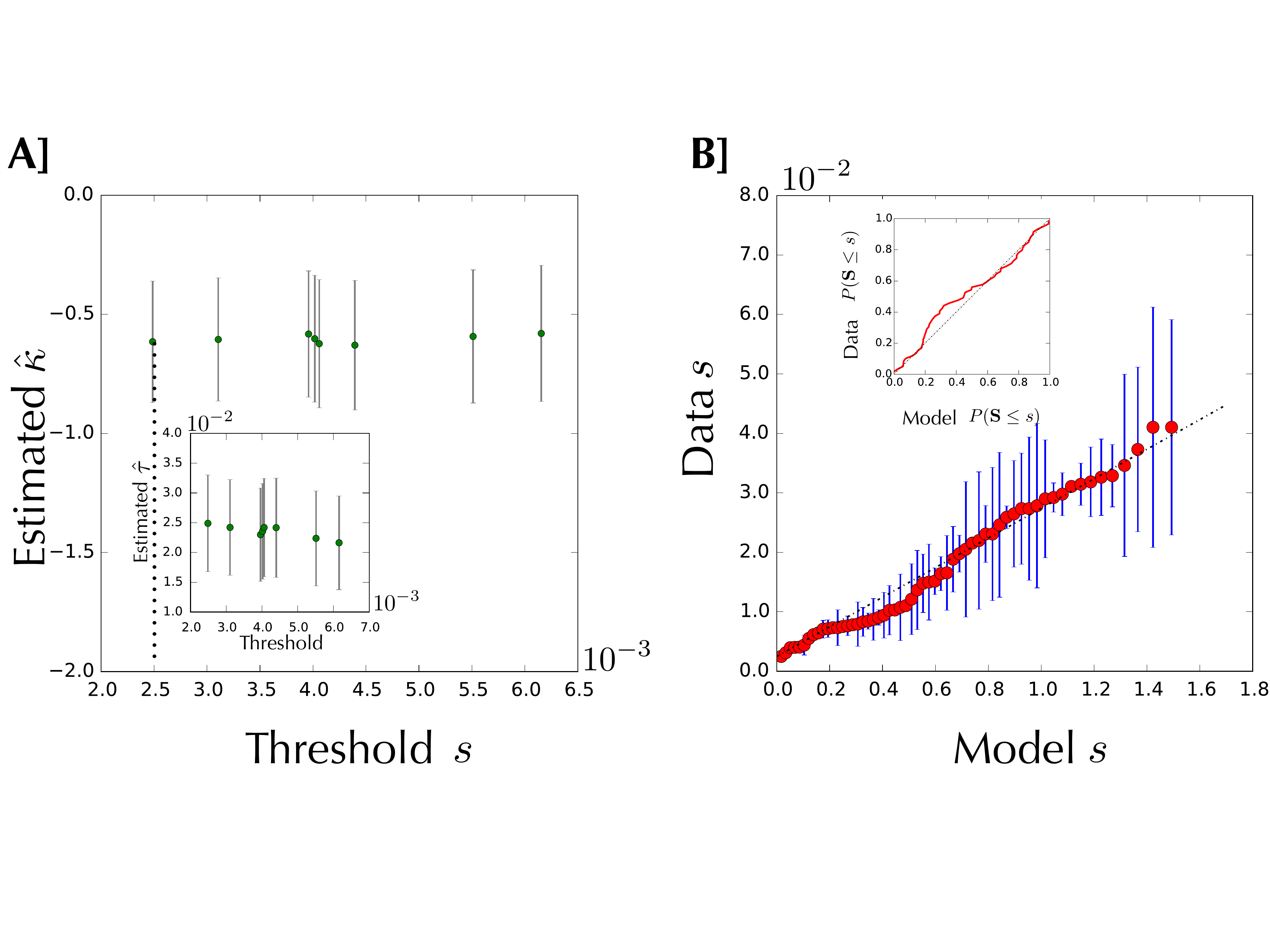}}
\caption{EVT analysis for the selection of the CH1 library against the DNA target (data shown in Figure 3C). A fit of the general model gives $\kappa=-0.62\pm 0.25$, $\tau=2.5\times10^{-2}\pm 8\times10^{-3}$ while a fit of the exponential model ($\k=0$) gives $\tau_0=1.5\times10^{-2}\pm 4\times10^{-3}$; the exponential model is excluded with a p-value $10^{-2}$, in favor of $\k<0$.\label{fig:chicken}}
\end{figure}

\begin{figure}[h]
\centerline{\includegraphics[width=.7\linewidth]{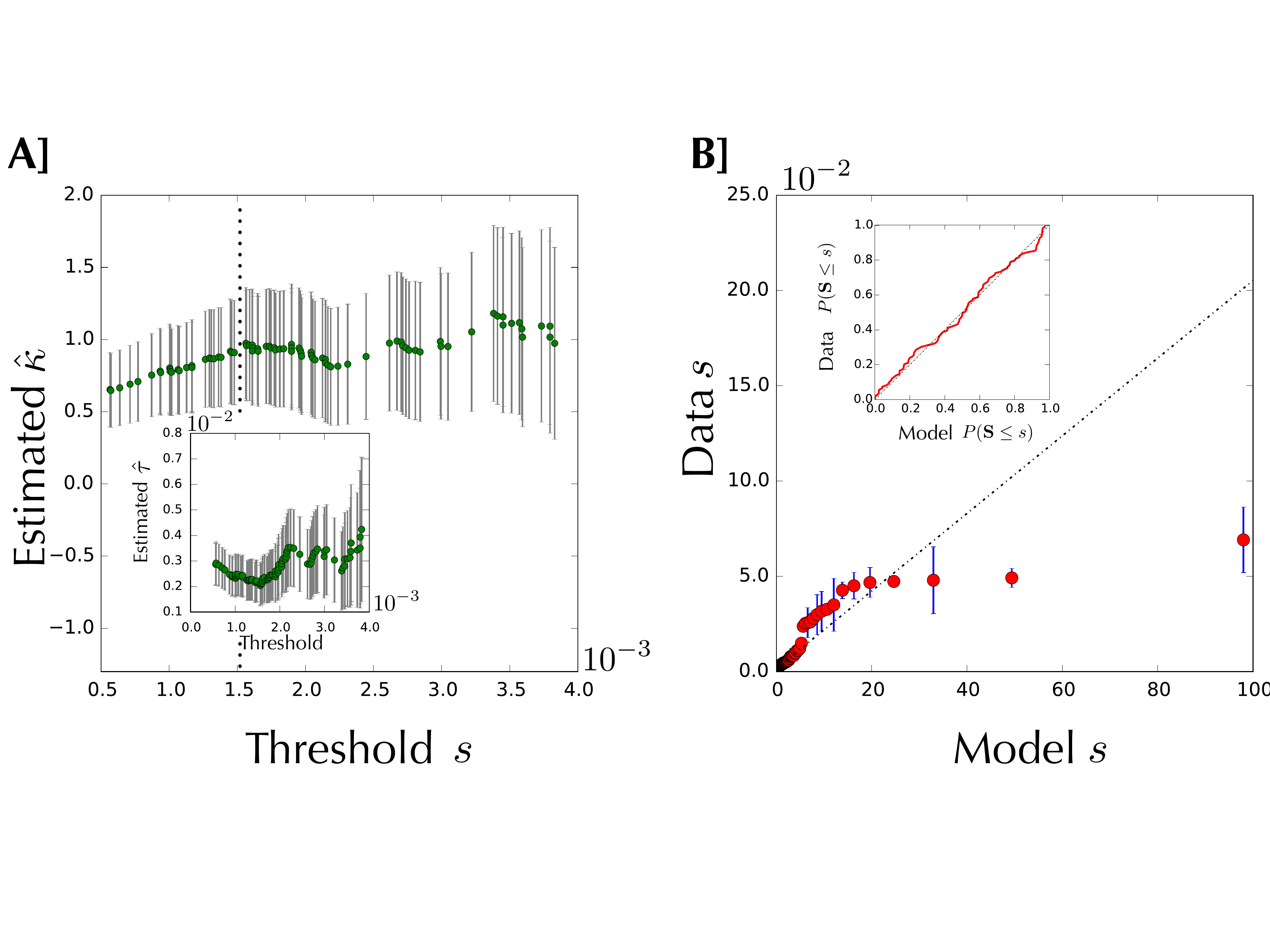}}
\caption{EVT analysis for the selection of the F3 library against the DNA target (data not shown in the main text). A fit of the general model gives $\kappa=0.97\pm 0.38$, $\tau=2\times10^{-3}\pm 8\times10^{-4}$ while a fit of the exponential model ($\k=0$) gives $\tau_0=7.5\times10^{-3}\pm 10^{-3}$; the exponential model is excluded with a p-value $<10^{-4}$, in favor of $\k>0$.\label{fig:Frog3noire}}
\end{figure}

\begin{figure}[h]
\centerline{\includegraphics[width=.7\linewidth]{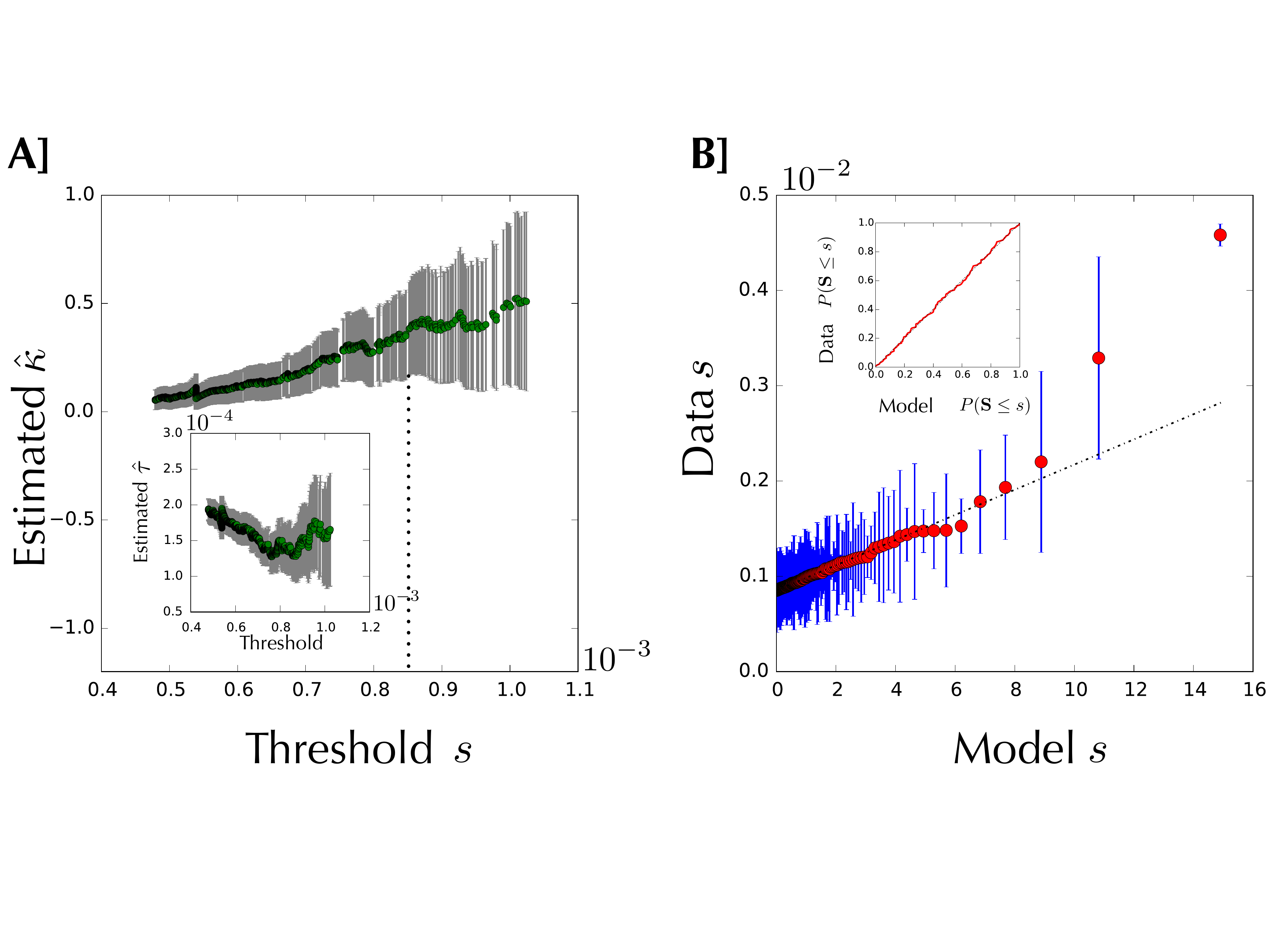}}
\caption{EVT analysis for the selection of the N1 library in the mixture of 24 libraries against the PVP target (data not shown in the main text). A fit of the general model gives $\kappa=0.38\pm 0.21$, $\tau=1.3\times10^{-4}\pm 3\times10^{-5}$ while a fit of the exponential model ($\k=0$) gives $\tau_0=2.2\times10^{-4}\pm 3\times10^{-5}$; the exponential model is excluded with a p-value $<10^{-4}$, in favor of $\k>0$. \label{fig:NursesharkMix24}}
\end{figure}

\begin{figure}[h]
\centerline{\includegraphics[width=.4\linewidth]{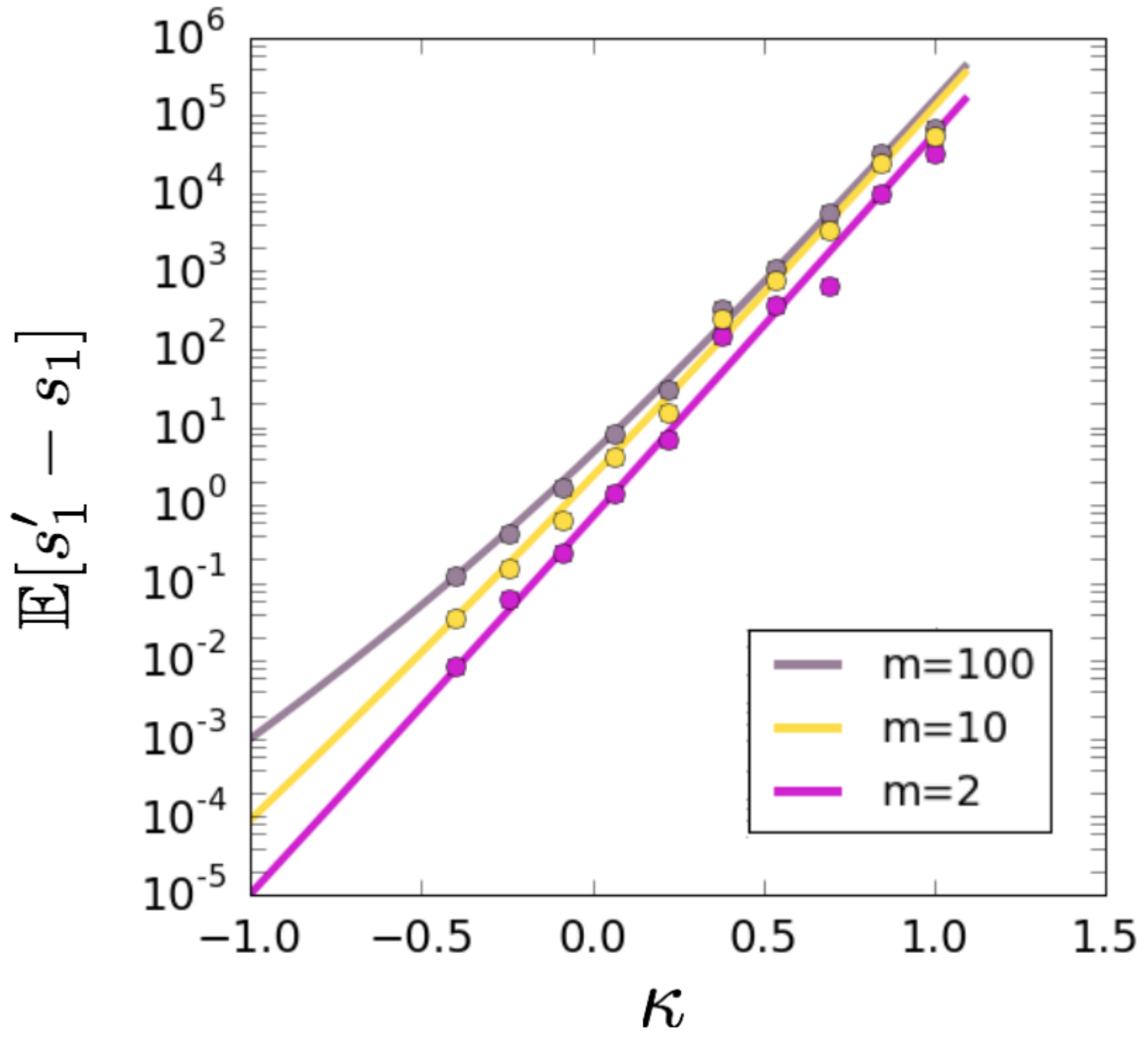}}
\caption{Scaling of the best binder with the library size -- To estimate the gain that sampling $m$ times more the same library may provide as a function of the shape parameter $\k$, we show here the expected difference $\E[s'_1-s_1]$ between the maximum $s_1'$ of $mN$ samples drawn with probability density $f_\k(x)$ from Eq.~(4) and the maximum $s_1$ of $N$ sub-samples. The plain lines are based on Eq.~\eqref{eq:scaling} with $\tau=1$ and $N=10^5$ and the dots are the results of numerical simulations (averaged over many draws), showing a good agreement between the two. Note how $\E[s'_1-s_1]$ depends more strongly on $\k$ than on $m$.\label{fig:scaling}}
\end{figure}

\begin{figure}[h]
\centerline{\includegraphics[width=.8\linewidth]{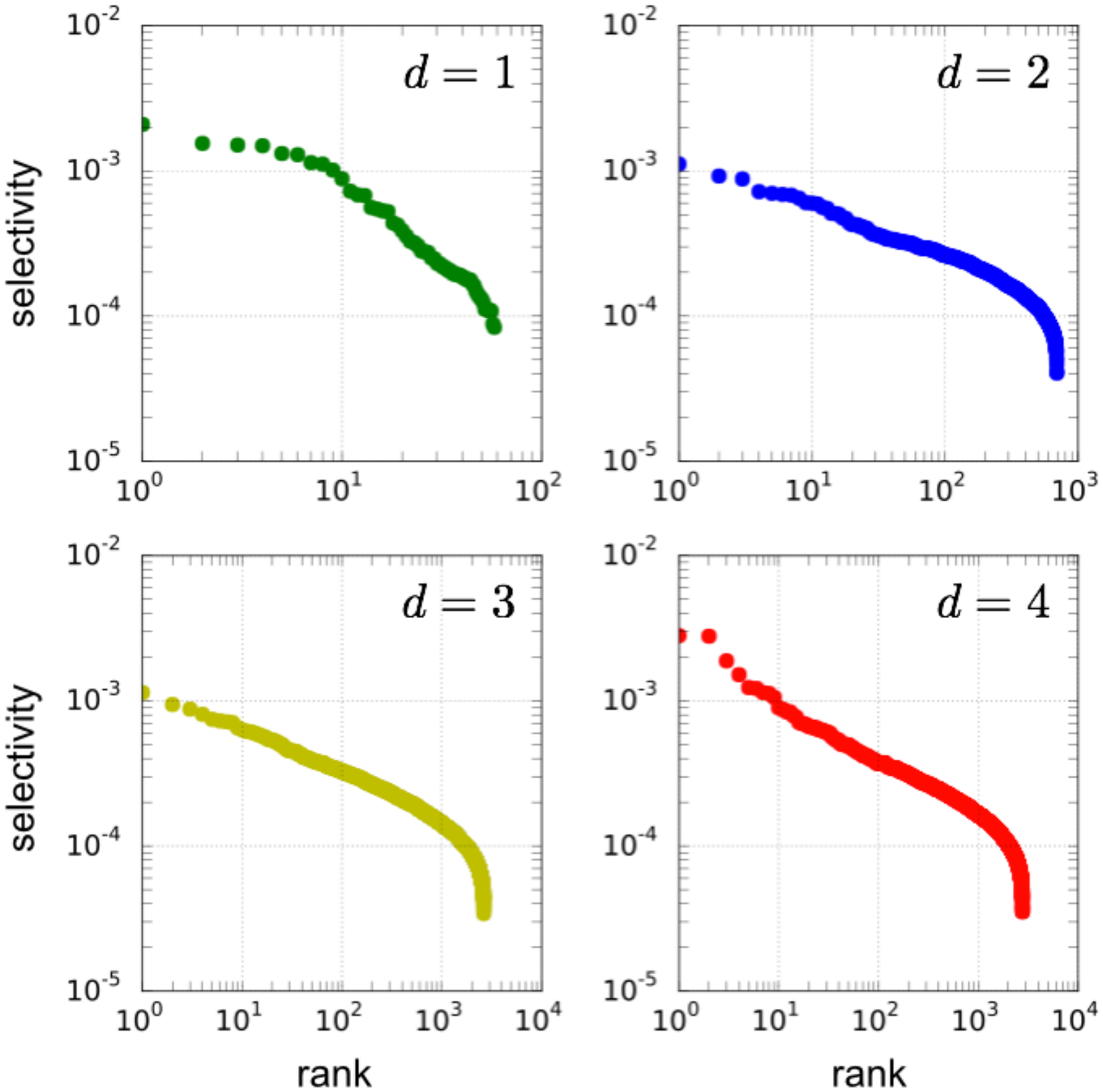}}
\caption{Stability of the shape parameter $\kappa$ for non-random sub-samples of a same library -- To test whether non-random sub-libraries may be expected to be described by the same shape parameter as the library from which they originate, we consider here the results of the selection of library S1 against PVP (Figure~3A), for which the consensus CDR3 has amino acids sequence GWYT and we make four non-overlapping sub-libraries consisting of sequences with CDR3 at distance $d=1$ to $4$ from this consensus, where the distance just counts the number of amino acid differences (number of mutations). This figure shows the selectivity versus the rank of the sequences in these sub-libraries. An EVT analysis indicates that $\k(d=1)=0.33\pm 0.39$, $\k(d=2)=0.40\pm0.26$, $\k(d=3)=0.30\pm 0.23$, $\k(d=4)=0.53\pm 0.22$: all these values are comparable to the value $\k=0.44\pm 0.22$ of the shape parameter for the full library.\label{fig:sub}}
\end{figure}

\begin{figure}[h]
\centerline{\includegraphics[width=.6\linewidth]{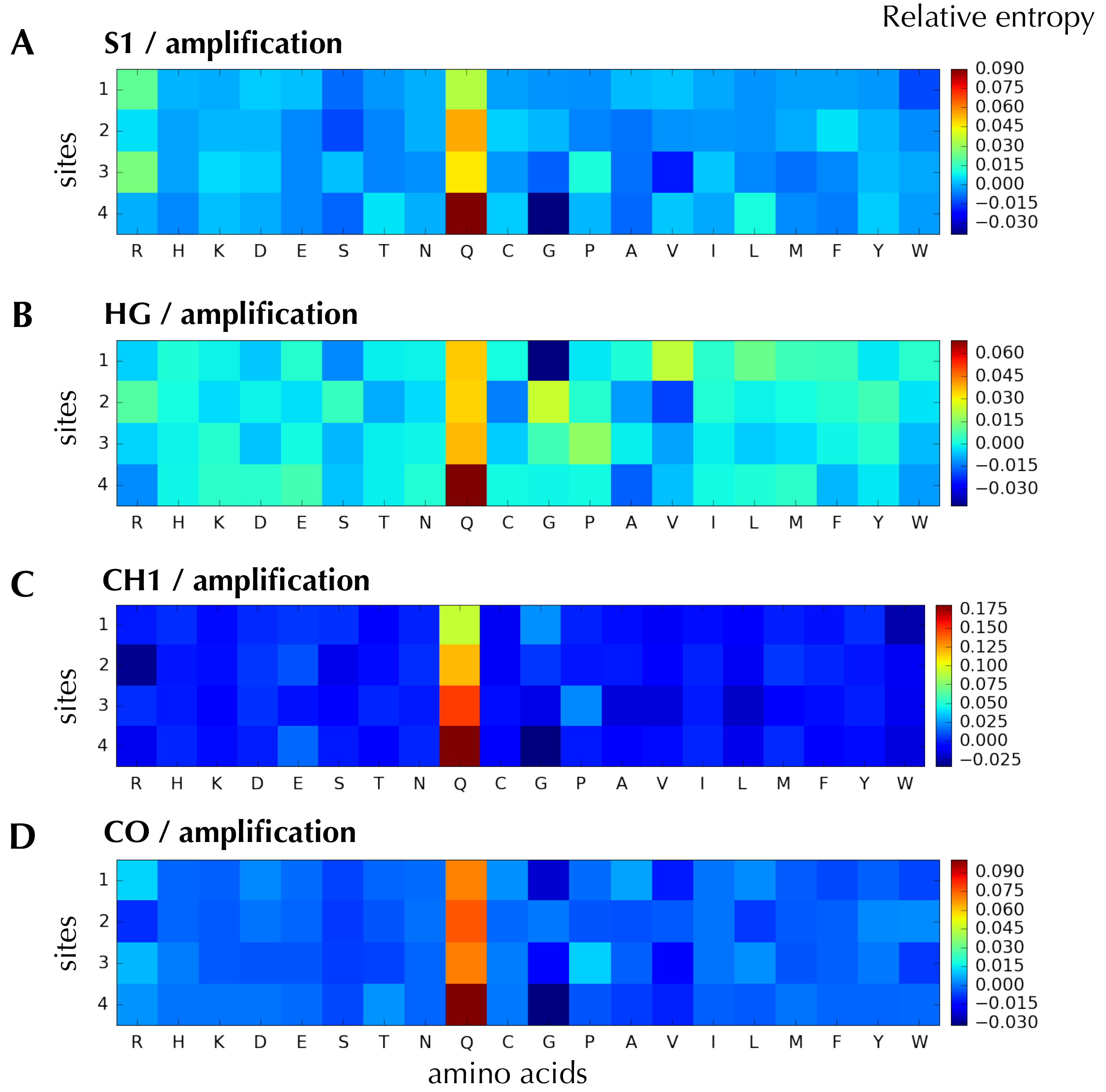}}
\caption{Amplification bias -- Relative entropy between frequencies of CDR3 sequences before and after amplification without selection, showing an enrichment in glutamine (represented by the letter Q). The results presented in the paper exclude sequences with an amber codon, which is responsible for this effect (see supplementary experimental methods), but, in most experiments with selection, glutamine does not appear in the selected consensus sequence and considering the amber code as coding for an amino acid or for a stop codon has no incidence on the conclusions. \label{fig:amp2}}
\end{figure}

\begin{figure}[h]
\centerline{\includegraphics[width=.9\linewidth]{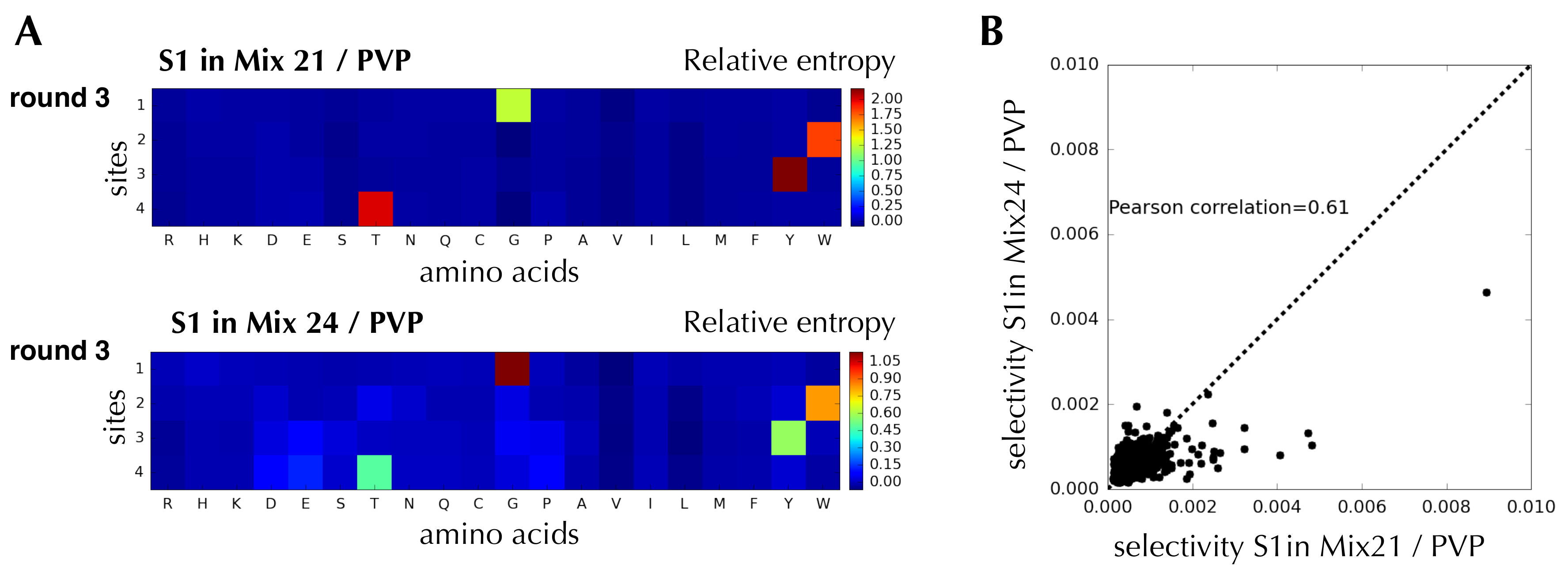}}
\caption{Reproducibility of selections against the PVP target -- The results of two experiments of selection against the PVP target, one starting from a mixture of 24 libraries and the other from a subset of 21 libraries, which each are dominated by the S1 library, not only lead to an identical consensus sequence (panel A) but to reproducible results by EVT analysis (Table S\ref{tab:robust}). In this case, not only are the initial populations different, but also potentially the targets since the experiments were performed 1.5 year apart and PVP is subject to aging: this may explain the imperfect correlations between frequencies (panel B; by contrast, the selection of the F3 library against PVP was performed at the same time than the selection of the mixture of 24 libraries and differences in consensus sequences cannot be due to differences of the targets in this case).\label{fig:reproNurse}}
\end{figure}

\begin{figure}[h]
\centerline{\includegraphics[width=.75\linewidth]{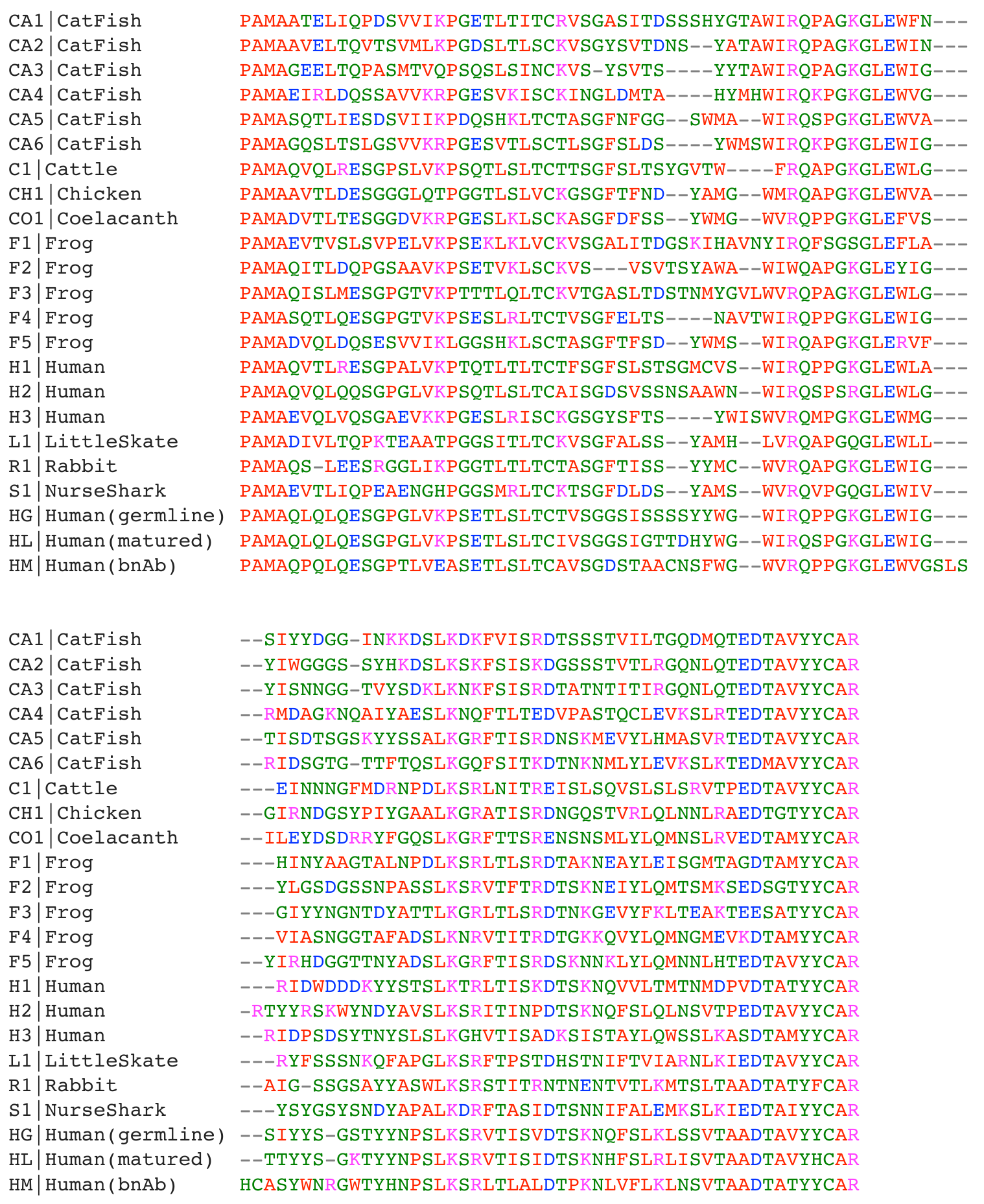}}
\caption{Amino acid sequences of the frameworks -- Multiple sequence alignment of the library-specific part of the frameworks (Figure~1). The organism from which the sequence originate is indicated.\label{fig:seq_fram}}
\end{figure}

\end{document}